\documentclass[aps,twocolumn,preprintnumbers,showpacs]{revtex4}
\usepackage{graphics,graphicx,color}	
\usepackage{amsmath}

\newcommand{\beq}{\begin{equation}}
\newcommand{\eeq}{\end  {equation}}

\newcommand{\beqar}{\begin{eqnarray}}
\newcommand{\eeqar}{\end  {eqnarray}}

\newcommand{\bfig}{\begin{figure}}
\newcommand{\efig}{\end  {figure}}

\newcommand{\bold}[1]{\mbox{\boldmath $#1$}}    
\newcommand{\GeV}{{\rm GeV}}                    
\newcommand{\MeV}{{\rm MeV}}                    
\newcommand{\fm}{{\rm fm}}                      
\newcommand{\rme}{{\rm e}}                      
\newcommand{\Z}{{\cal Z}}

\newcommand{\half}{\mbox{${1\over2}$}}          
\newcommand{\third}{\mbox{${1\over3}$}}         
\newcommand{\quart}{\mbox{${1\over4}$}}         
\newcommand{\sixth}{\mbox{${1\over6}$}}         

\newcommand{\ZERO}[1]{~\hspace{0.4ex}
  \raisebox{1.7ex}{\scriptsize {$\circ$}}\hspace{-1.6ex}{#1}}

\begin{document}


\preprint{LBNL-57871}

\title{Signals of spinodal hadronization: strangeness trapping}

\author{Volker Koch, Abhijit Majumder, and J{\o}rgen Randrup}

\affiliation{Nuclear Science Division, Lawrence Berkeley National Laboratory,
1 Cyclotron Road, Berkeley, CA 94720}

\date{\today}

\begin{abstract}
If the deconfinement phase transformation of strongly interacting matter
is of first-order and 
the expanding chromodynamic matter created in a
high-energy nuclear collision enters the corresponding
region of phase coexistence,
a spinodal phase separation might occur.
The matter would then condense into a number of separate blobs,
each having a particular net strangeness
that would remain approximately conserved during the further evolution.
We investigate the effect that such {\em strangeness trapping}
may have on strangeness-related hadronic observables.
The kaon multiplicity fluctuations are significantly enhanced
and thus provide a possible tool 
for probing the nature of the phase transition experimentally.
\end{abstract}


\maketitle


\section{Introduction}

One of the major goals of high-energy heavy-ion research 
is to explore the equation of state of strongly interacting matter,
particularly its phase structure \cite{goals}.
Depending on the beam energy,
various regions of temperature and baryon density can be explored.
Thus systems with a very small net baryon density 
are formed at RHIC \cite{Adler:2001bp} $(\sqrt{s} \simeq 200~A\GeV)$,
while it is expected that the creation of the 
highest possible baryon densities occurs at
more moderate beam energies $(\sqrt{s} \simeq 10~A\GeV)$,
such as those becoming available at the planned FAIR \cite{FAIR}.

Our understanding of the QCD phase diagram is best developed
at vanishing chemical potential, $\mu_B=0$,
where lattice QCD calculations are most easily carried out.
The most recent results indicate that the transformation
from a low-entropy hadron resonance gas
to a high-entropy quark-gluon plasma
occurs smoothly at the temperature is raised,
with no real phase transition being present \cite{karsch_qm}.
On the other hand, at zero temperature
most models predict the occurrence of a first-order phase transition 
when the density is raised \cite{Stephanov:1998dy},
though no firm results are yet available
for the corresponding value of the chemical potential, $\mu_0$.
However, if the $T=0$ transformation is in fact of first order,
one would expect the phase boundary 
to extend into the region of finite temperature and terminate 
at a certain critical endpoint, $(\mu_c,T_c)$ \cite{Stephanov:1998dy}.
Indeed, recent lattice QCD results \cite{fodor} suggest the presence 
of such a first-order phase transition line
and an associated critical end-point,
though its precise location is not well determined.

It is therefore important to consider how this key issue
could be elucidated on the basis of experimental data.
Generally, one might expect that if the expanding matter created 
in a high-energy nuclear collision 
crosses a first-order phase-transition line
then the associated non-monotonic behavior of the thermodynamic potential
might have observational consequences.

From this perspective,
the enhancements of the $K/\pi$ ratio reported for
beam energies of $20-30\,A\GeV$ at the SPS \cite{NA49_kp} appears intriguing.
Since these data present the {\em only} non-monotonic behavior
seen so far in high-energy heavy-ion collisions,
it appears appropriate and timely to study the consequences 
of a possible first-order phase transition on the production of
kaons or, more generally, strange hadrons.

A universal feature of first-order phase transitions
is the occurrence of spinodal decomposition,
which results from the convex anomaly
in the associated thermodynamic potential \cite{PhysRep}.
This phenomenon occurs when bulk matter, by a sudden expansion or cooling,
is brought into the convex region of phase coexistence.
Since such a configuration is thermodynamically unfavorable
and mechanically unstable,
the uniform system seeks to reorganize itself into spatially separate
single-phase domains.
Moreover, since this spinodal phase separation develops by means of
the most unstable collective modes, the resulting domain pattern tends 
to have a scale characteristic of those modes.
This general phenomenon, which is known in many areas of physics
and has found a variety of technological applications,
appears to be an important mechanism behind the multifragmentation
phenomenon in medium-energy nuclear collisions \cite{INDRA,PhysRep},
where the relevant first-order phase-transition
is between the nuclear liquid and a gas of nucleons and light fragments.
Thus, if a first-order phase transition is encountered
during the expansion stage of a high-energy nuclear collision,
one might expect that such a spinodal separation might occur.
While the resulting enhancement of baryon fluctuations was studied by
Bower and Gavin \cite{bower} and
the prospects for observing such a process via $N$-body kinematic correlations 
was discussed by Randrup \cite{JR:HIP},
the present study explores the consequences
for the production of strange hadrons.

In order for a spinodal decomposition to occur,
several conditions must be met.
First of all, of course, the equation of state must have a first-order
phase transition.
Though expected, the existence of a first-order phase transition
is not yet well established theoretically 
and it may ultimately have to be determined experimentally
by analyzes of the kind considered here.
Second, the dynamical trajectory of the bulk matter formed early on
must pass through the spinodal region of phase coexistence.
While some calculations suggest that this may happen at FAIR 
\cite{Toneev,Weber:1998zb,Cassing:2000bj},
this question needs to be investigated more thoroughly.
Third, even if the the above conditions are met,
the dynamical conditions of the collision
must be carefully tuned to ensure, on the one hand,
that the bulk of the system is brought into the spinodal region 
sufficiently quickly to achieve a quench,
yet, on the other hand, the overall expansion should be slowed down
to a degree that will allow the dominant instabilities to grow sufficiently
to cause the bulk to break up.
While the conditions for achieving this delicate balance
are hard to ascertain theoretically,
they may be found by a systematic variation of the beam energy. 
These open questions notwithstanding,
we shall here assume that the matter created in a heavy-ion collision
somehow breaks into a number of subsystems, blobs,
which subsequently expand and hadronize independently
and we then investigate the consequences for the production of strange hadrons.
In particular, we wish to ascertain whether such a breakup 
could lead to an enhancement of the $K/\pi$  ratio and its fluctuations.

In such a scenario, if the breakup is sufficiently rapid, 
then whatever net strangeness happens to reside within the region of the plasma
that forms a given blob will effectively become trapped and, consequently,
the resulting hadronization of the blob 
will be subject to a corresponding constraint on the net strangeness.
As we shall demonstrate, this type of canonical constraint
will enhance the multiplicity of strangeness-carrying hadrons,
as compared to the conventional (grand-canonical) scenario where 
global chemical equilibrium is maintained through the hadronic freeze-out
\cite{PBM}.
(This occurrence of an enhancement is qualitatively easy to understand,
since the presence of a finite amount of strangeness in the hadronizing blob
enforces the production of a corresponding minimum number of strange hadrons.)
The fluctuations in the multiplicity of strange hadrons, such as kaons, 
are enhanced even more, thus offering a possible means
for the experimental exploration of the phenomenon.

The remainder of this paper is organized as follows:
First we introduce a suitable idealized model framework and
develop the necessary formal tools for the required canonical calculations,
with some of the formal manipulations being relegated to appendices.
The key features are then brought out in a schematic scenario
containing only charged kaons.
Subsequently, we present instructive numerical results
and also make a quantitative assessment of the importance
of global strangeness conservation.
We finally give a concluding discussion.

\section{Calculational framework}
\label{sec:calc_frame}

In order to establish a framework for investigating the effect
of the strangeness trapping mechanism,
we adopt the following schematic scenario:
We start by considering the expanding system 
when it is still in the plasma phase.
At this stage the system is spatially uniform
and the strange quarks and antiquarks can be considered
as being randomly distributed throughout the system,
irrespective of what the net baryon density happens to be.
We imagine that the bulk of the expanding and cooling plasma
enters the region of phase coexistence and that the associated spinodal 
instability will cause it to break up into separate subsystems, blobs,
which are assumed to all have the same size,
as they would tend to have in a spinodal breakup.
Each of these blobs now proceed to expand and hadronize
while maintaining its net strangeness.
The resulting assembly of hadrons is determined at freeze-out
by a sampling of the statistical phase space,
subject to the appropriate canonical strangeness constraint.

In order to assess the effect of the strangeness trapping,
it is useful to compare the results against the standard scenario, 
in which the system is assumed to evolve to freeze-out 
while remaining macroscopically uniform and
maintaining global statistical equilibrium.
Focusing on a particular subvolume $V_h$,
we describe the resulting hadron gas in the classical 
grand-canonical approximation.
The abundance of a particular hadron specie $k$ is then
\beq
\bar{n}_k = {g_k\over2\pi^2}\, {V_hT_h^3\over\hbar^3c^3}\,
\tilde{K}_2({m_kc^2\over T_h})\, \rme^{(\mu_BB_k+\mu_QQ_k+\mu_SS_k)/T_h} ,
\eeq
where $\tilde{K}_2(x)\equiv x^2K_2(x)$ is regular at $x=0$ and
a hadron of the specie $k$ has baryon number $B_k$, electric charge $Q_k$,
and strangeness $S_k$.
The average values of baryon number, charge, and strangeness
in the volume $V_h$ then readily follow,
\beq\label{BQS}
\bar{B}=\sum_k B_k\bar{n}_k\ ,\
\bar{Q}=\sum_k Q_k\bar{n}_k\ ,\
\bar{S}=\sum_k S_k\bar{n}_k\ .
\eeq
The values of the freeze-out temperature $T_h$ 
and the three chemical potentials $\mu_B$, $\mu_Q$, and $\mu_S$
will be determined by fits to the experimental yield ratios (see later).
In this treatment, the individual hadron species are statistically independent
and the associated multiplicities have Poisson distributions,
so the multiplicity variances are equal to the mean values,
$\sigma_k^2=\bar{n}_k$.

We now return to the particular spinodal scenario described above,
where we assume that the plasma has broken up into separate blobs.
We first consider the distribution of strangeness within the blobs
and then treat their subsequent hadronic freeze-out.

If a given plasma blob is only a small part of the total system,
its statistical properties may be treated in the grand-canonical approximation.
The various quark (and gluon) species are then independent.
Furthermore, since there is no bias on the overall strangeness,
the $s$ and $\bar s$ quarks have identical partition functions,
${\cal Z}_s={\cal Z}_{\bar s}$, where
\beqar
\ln{\cal Z}_s &=&
g_q \int{d^3\bold{r}_s d^3\bold{p}_s\over h^3}\,
\ln[1+\rme^{-\epsilon_s/T_q}]\\ 
&\approx& {3\over\pi^2} {V_q T_q^3\over\hbar^3c^3} 
\tilde{K}_2\left({m_s\over T_q}\right)\ \equiv\ \zeta_s\ .
\eeqar
Here $g_q=6$ is the quark spin-color degeneracy,
$T_q$ is the plasma temperature, 
and $V_q$ is the volume of the particular blob considered
at the time when its strangeness is frozen in.
The energy $\epsilon_s$ is given by $\epsilon_s^2=p_s^2+m_s^2$,
where we use the mass $m_s=150\,\MeV$.
Whereas the first expression is the exact fermionic form,
the last relation emerges in the classical limit
which we shall adopt here for simplicity (see Appendix \ref{fermions}).
Then the $s$ and $\bar s$ multiplicities, $\nu$ and $\bar{\nu}$,
have Poisson distributions characterized by the mean value
$\zeta_s=\ln{\cal Z}_s$.
The total strangeness content in the blob is then $S_0=\bar{\nu}-\nu$.
Since the different quark flavors are distributed independently,
the resulting probability for ending up with a given blob strangeness $S_0$
is independent of the prevailing baryon and charge contents
and can be expressed as a modified Bessel function,
\beq\label{prob_S0}
P(S_0) = \sum_{\nu\bar\nu}
{\zeta_s^\nu\zeta_s^{\bar\nu}\over{\nu!\bar\nu}!}\,\rme^{-2\zeta_s}\ 
\delta_{\bar{\nu}-\nu,S_0} = I_{S_0}(2\zeta_s)\,\rme^{-2\zeta_s}\ . 
\eeq
We note that the corresponding ensemble average value of $S_0$ vanishes, 
$\prec\! S_0\!\succ=0$,
while its ensemble variance is 
$\sigma_{S_0}^2=\sum_{S_0} S_0^2\, P(S_0)=2\zeta_s$.

To avoid spurious correlations, we take account of the fact that
the presence of a certain net strangeness in a given blob
introduces a bias on its baryon number and charge,
relative to the overall grand-canonical averages $\bar B$ and $\bar Q$
of the grand-canonical reference scenario, Eq.\ (\ref{BQS}).
Indeed, each particular value of $S_0$ determines a canonical subensemble 
of blobs that have modified distributions of baryon number and charge.
It is elementary to show that
these are shifted by amounts proportional to $S_0$,
so the corresponding conditional expectation values become
\beq\label{BQ}
\langle B\rangle_{S_0}=\bar{B}-\third S_0\ ,\,\
\langle Q\rangle_{S_0}=\bar{Q}+\third S_0\ .
\eeq
It also follows that the ensemble correlations of $B$ and $Q$ 
with $S$ are given by
\beqar\label{BQ-S}
\sigma_{BS} &=& \prec B S\succ\ =\ -\third \sigma_{S_0}^2\ ,\\
\sigma_{QS} &=& \prec Q S\succ\ =\ +\third \sigma_{S_0}^2\ .
\eeqar

Each isolated blob is assumed to expand further while hadronizing,
until the freezeout volume $V_h=\chi V_q$ has been reached.
Its temperature is then $T_h\leq T_q$.
Rough approximations to the equation of state \cite{RandrupPRL92}
and the demand of energy conservation
suggest that the expansion factor is $\chi\approx3$,
to within a factor of two or so.
The blob has now transformed itself into a hadron resonance gas
which we describe as a canonical ensemble
characterized by the strangeness $S_0$ of the precursor plasma blob.
The baryon number and charge are treated grand-canonically
and the demand that the expected baryon and charge contents 
match the above conditional values (\ref{BQ}) then determines the associated
{\em biased} chemical potentials $\mu_B'$ and $\mu_Q'$,
where the primes indicate that these pertain to the biased
(canonical) ensemble characterized by the particular value of $S_0$.
Though required for formal consistency
and included in the calculations,
this refinement is not quantitatively important.

For the description of the hadron gas,
we include 124 hadronic species $\{k\}$, 
from the $\pi^0(135)$ and up though the $\Omega^-(1672)$.
Each specie is characterized by its one-particle partition function,
\beqar
\zeta_k &=& g_k \int{d^3\bold{r}_k d^3\bold{p}_k\over h^3}\,
\rme^{-(\epsilon_k-\mu_B'B_k-\mu_Q'Q_k)/T_h}\\ \label{mu'}
&=& {g_k\over2\pi^2}{V_hT_h^3\over\hbar^3c^3}\tilde{K}_2({m_kc^2\over T_h})\,
 \rme^{(\mu_B'B_k+\mu_Q'Q_k)/T_h}\ . \label{zeta_k}
\eeqar
The non-strange hadrons,
which are not affected directly by the canonical strangeness constraint,
have grand-canonical distributions governed by 
the {biased} chemical potentials $\mu_B'(S_0)$ and $\mu_Q'(S_0)$.
The total partition function is therefore of the form
${\cal Z}={\cal Z}_{\rm gc}^{\{S=0\}}{\cal Z}_{S_0}^{\{S\neq0\}}$.
We describe briefly below how we obtain the canonical partition function
for the strange hadrons, ${\cal Z}_{S_0}^{\{S\neq0\}}$,
and refer to Appendix \ref{formulas} for more details.

For this task, we organize the hadron species according to their strangeness
(an alternate approach was employed in 
Refs.\ \cite{Majumder:2003gr,Cleymans:2004iu}).
For each strangeness class $S=\pm1,\pm2,\pm3$,
we introduce the {\em effective} one-particle partition function 
$\zeta_S=\sum_\kappa\zeta_\kappa\delta_{S_\kappa,S}$,
which accounts for all hadron species having the particular strangeness $S$.
The corresponding {\em generic} multiplicities are denoted by $N_S$,
the multiplicity of hadrons having the specified strangeness $S$
({\em e.g.}\ $N_{+1}=n_{K^0}+n_{K^+}+n_{\bar\Lambda}+\dots$).
It should be noted that $\zeta_S$ is not simply related to $\zeta_{-S}$ 
when the system has a net baryon density.
In terms of these quantities we then have
\beq\label{canonical_part}
{\cal Z}_{S_0}^{\{S\neq0\}}\ =\ \prod_{S\neq0}\left[\sum_{N_S\geq0} 
{\zeta_S^{N_S}\over N_S!}\right]\, \delta(\sum_S N_SS-S_0)\ . 
\eeq

It is convenient to obtain this total partition function recursively
by adding two conjugate strangeness classes $\{\pm S\}$ at a time,
\beqar
&~&\!\! {\cal Z}_{S_0}^{\{S=\pm1,\pm2\}} =
\sum_{S_2=0,\pm2,\dots}{\cal Z}_{S_0-S_2}^{\{\pm1\}}\ 
{\cal Z}_{S_2}^{\{\pm2\}} ,\\
&~&\!\! {\cal Z}_{S_0}^{\{S=\pm1,\pm2,\pm3\}} =
\sum_{S_3=0,\pm3,\dots}{\cal Z}_{S_0-S_3}^{\{\pm1,\pm2\}}\ 
{\cal Z}_{S_3}^{\{\pm3\}} ,\
\eeqar
where the canonical partition function for a single pair of conjugate 
strangeness classes is given by an expression analogous to
 Eq.\ (\ref{prob_S0}) \cite{Cleymans:2004iu},
\beqar
{\cal Z}_{S_0}^{\{\pm S\}} &=& \sum_{N_+N_-}
{\zeta_{+S}^{N_+}\over N_+!}\, {\zeta_{-S}^{N_-}\over N_-!}\, 
\delta_{(N_+-N_-)S,S_0}\ \nonumber \\
&=& \left({\zeta_{+S}\over\zeta_{-S}}\right)^{S_0/2}
I_{S_0}\left(2\sqrt{\zeta_{+S}\zeta_{-S}}\right)\ .
\eeqar
The corresponding correlated conditional distribution 
of the generic multiplicities $\{N_S\}$ is then determined.
Furthermore, recursive expressions can readily be derived
for the associated multiplicity moments,
such as $\langle N_S\rangle_{S_0}$ and $\langle N_S N_{S'}\rangle_{S_0}$
(see Appendix \ref{formulas}).
Once we know the number of hadrons with a given strangeness $S$, $N_S$,
we can obtain the multiplicities $\{n_\kappa\}$ of the individual strange 
species from the corresponding binomial distributions,
\beq\label{Pn}
P_\kappa(n_\kappa)\ =\ 
{N_S!\over\zeta_S^{N_S}}{\zeta_\kappa^{n_\kappa}\over n_\kappa!}
{(\zeta_S-\zeta_\kappa)^{N_S-n_\kappa}\over(N_S-n_\kappa)!}\ .
\eeq

By proceeding as described above,
it is possible to treat the entire ensemble of possible blobs
and by numerical simulation generate a sample of ``events''
consisting of the resulting primordial hadrons.
(The importance of subsequent decays is discussed later.)
These ``final states'' can then be analyzed as idealized experimental data.
This can be done both in the spinodal scenario
where individual blobs are treated canonically
as well as in the grand-canonical reference scenario.
For instructive purposes, we also consider a restricted canonical scenario
in which the strangeness of a blob is always required to vanish, $S_0=0$.

The inclusion of the bias effect (see Eq.\ (\ref{BQ})), 
as expressed through $\mu_B'$ and $\mu_Q'$ in (\ref{mu'}),
ensures that the ensemble correlation of baryon number and charge 
with strangeness remains the same as it were in the plasma,
even though our treatment of the hadron production
conserves $B$ and $Q$ only on the average for each $S_0$,
\beqar\nonumber
\sigma_{BS} &\equiv& \prec BS\succ-\prec B\succ\prec S\succ\
=\ \sum_{S_0}\langle B\rangle_{S_0} S_0 P(S_0)\\
&=& \sum_{S_0}(\bar{B}-\third S_0) S_0 P(S_0)\
=\ -\third\sigma_{S_0}^2\ ,\\ \nonumber
\sigma_{QS} &\equiv& \prec QS\succ-\prec Q\succ\prec S\succ\
=\ \sum_{S_0}\langle Q\rangle_{S_0} S_0 P(S_0)\\
&=& \sum_{S_0}(\bar{Q}+\third S_0) S_0 P(S_0)\
=\ +\third\sigma_{S_0}^2\ .
\eeqar

\section{Schematic scenario}
\label{sec:schematic-model}

Before discussing the results of such numerical simulations,
it is instructive to first illustrate the effect of strangeness trapping
in a schematic scenario.
For this purpose, assume that plasma blobs of strangeness $S_0$ hadronize 
into charged kaons only. We furthermore ignore the chemical potentials
($\mu_B$ is ineffective for mesons and $\mu_Q$ is anyway rather small).
The resulting $K^\pm$ multiplicity distributions are then given in terms of
the (common) one-particle partition function,
\begin{equation}
\zeta_K = g_K\!\! \int{d^3\bold{r} d^3\bold{p} \over h^3}\,
\rme^{-\epsilon_K/T_0} =
{g_K\over2\pi^2}{V_hT_h^3\over\hbar^3c^3}\tilde{K}_2({m_K\over T_h})\ .
\end{equation}
In particular, the average kaon multiplicities 
resulting from blobs of strangeness $S_0$ can be expressed as 
an asymptotic expansion in $1/\zeta_K$,
\begin{equation}
\langle n_{K^\pm}\rangle_{S_0}\ =\
\zeta_K \pm\half S_0-\quart +{4S_0^2-1\over32\zeta_K}\
+\ {\cal O}({1\over\zeta_K^2})\ .
\end{equation}
Ignoring the relatively small effects of quantum statistics
on the quark multiplicity distribution,
we found above that the ensemble of values $S_0=\bar{\nu}-\nu$
is characterized by $\prec S_0\succ = 0$ and $\sigma_{S_0}^2 = 2\zeta_s$,
The ensemble average $K^\pm$ multiplicities are then
\begin{equation}
\prec n_{K^\pm}\succ\ =\ \zeta_K -\quart +{8\zeta_s-1\over32\zeta_K}\
+\ {\cal O}({1\over\zeta_K^2})\ .
\end{equation}
Furthermore, the corresponding expressions 
for the ensemble multiplicity (co)variances are
\begin{eqnarray}
&~& \sigma_{K^+}^2\ =\ \sigma_{K^-}^2\ =\ 
	\half\zeta_K +\half\zeta_s +{\cal O}({1\over\zeta_K})\ ,\\
&~& \sigma_{K^+K^-}^2\ =\
	 \half\zeta_K -\half\zeta_s +{\cal O}({1\over\zeta_K})\ .
\end{eqnarray}
It should be noted that these expressions imply that
$\sigma^2(n_{K^+}\!-\!n_{K^-})=2\zeta_s=\sigma_{S_0}^2$
as required by the fact that $S_0=n_{K^+}-n_{K^-}$ in each blob.

In the special case where $\zeta_s$ happens to equal $\zeta_K$
we recover the grand-canonical result,
$\prec n_{K^\pm}\succ=\sigma_{K^\pm}^2=\zeta_K$ 
and $\sigma_{K^+K^-}=0$,
reflecting the fact that the grand-canonical ensemble can be built
from an ensemble of canonical subensembles.
But generally the above results differ from those
of the  grand-canonical scenario.
In particular, when $\zeta_K<\zeta_s=(1+\Delta)\zeta_K$,
as tends to be the case due to the large degeneracy of the quarks,
the kaon production is {\em enhanced} by the strangeness trapping,
\begin{eqnarray}
\prec n_{K^\pm}\succ &\approx& \zeta_K+\quart\Delta\ ,\\
\sigma_{n_{K^\pm}}^2 &\approx& \zeta_K+\half \Delta\zeta_K\ ,\\
\sigma_{n_{K^+}n_{K^-}} &\approx& -\half\Delta\zeta_K\ .
\end{eqnarray}
The negative value of the covariance reflects the fact
positive (negative) values of $S_0$ favors positive (negative) kaons,
so an excess of positive kaons is likely to be accompanied 
by a deficit of negative kaons, and vice versa.
We note that while the relative increase of the average multiplicities
is of the order of $\Delta/\zeta_K$,
the relative effect on the fluctuations is of the order of $\Delta$,
thus being enhanced by a factor of $\zeta_K$.
This basic feature suggests that the fluctuations are preferable
to the averages for probing the strangeness trapping phenomenon.

\section{Results and Discussion}
\label{sec:results}

We now discuss results obtained by numerical sampling
of the canonical partition function described above.
In our calculations, 
which serve to merely illustrate the effect
of the strangeness trapping mechanism,
we use the freeze-out hadron populations directly in the analysis
and make no attempt to include subsequent electroweak decays.
Obviously, this complication should be addressed
before a quantitative confrontation with data can be made.

\begin{figure}          
\includegraphics[angle=0,width=3.1in]{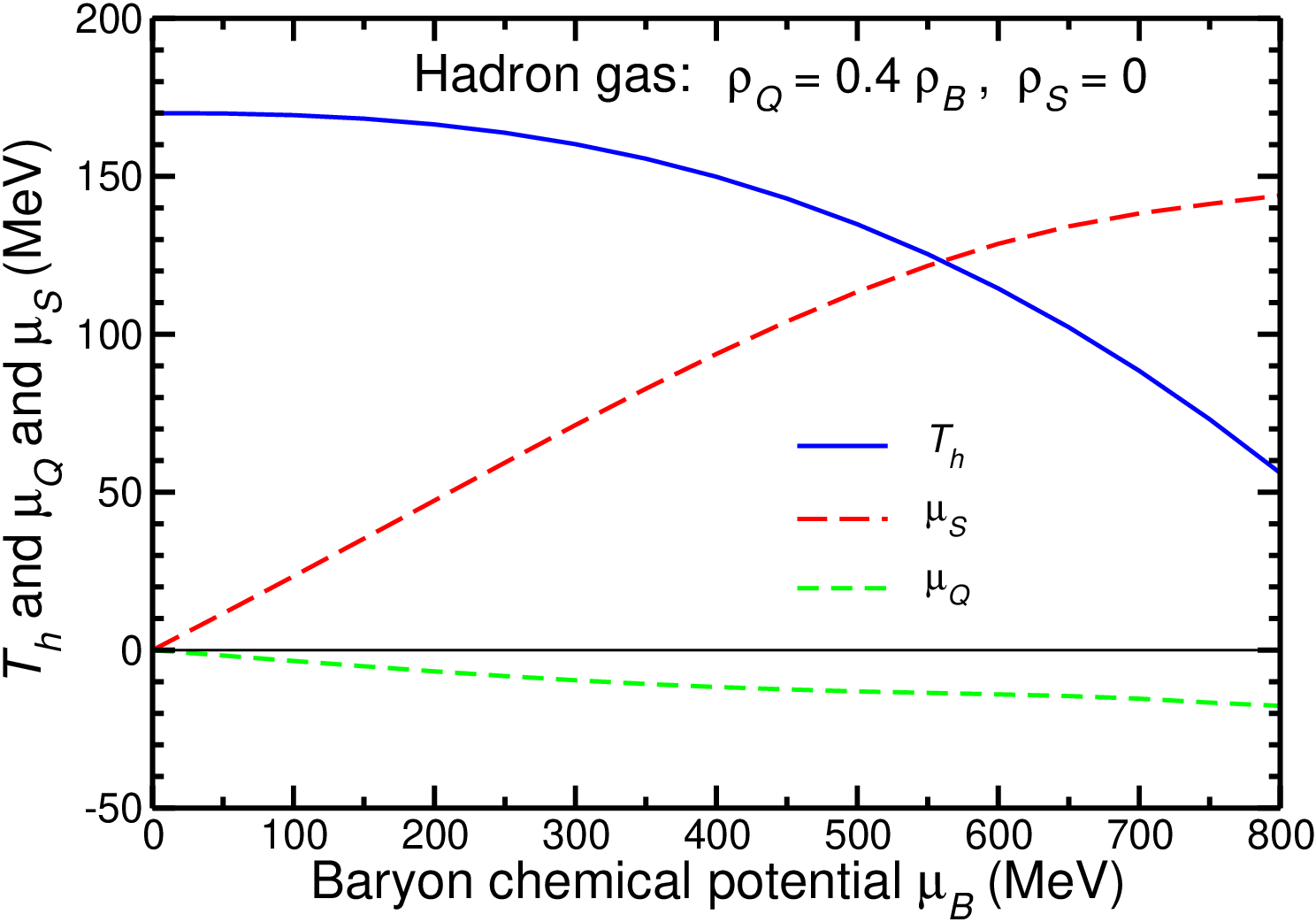}	
\caption{The employed freeze-out values of the temperature $T_h$
and the chemical potentials $\mu_Q$ and $\mu_S$ as functions of $\mu_B$,
as obtained by fitting to the result for $T_h(\mu_B)$ \cite{PBM}
and demanding that $\bar{Q}=0.4\bar{B}$ and $\bar{S}=0$.
}\label{f:TQS}
\end{figure}            

The overall grand-canonical reference environment is determined as follows.
Using the baryon chemical potential $\mu_B$ as a control parameter,
we obtain the freeze-out temperature $T_h$
from the fit to the data obtained in Ref.\ \cite{PBM},
yielding $T_h(\mu_B)\approx T_0[1-(\mu_B/m_N)^{5/2}]$ with $T_0=170~\MeV$.
Subsequently, we perform a grand-canonical iteration to determine 
those values of $\mu_Q$ and $\mu_S$ that ensure
$\bar{Q}=\alpha\bar{B}$ and $\bar{S}=0$,
where $\alpha=0.4$ which is representative of $Z/A$ for gold.
The resulting values are shown in Fig.\ \ref{f:TQS}
as functions of $\mu_B$.
As $\mu_B$ is increased, baryons become favored over antibaryons,
so there will a bias of hyperons over antihyperons.
To counterbalance the associated net negative strangeness 
(recall that $S_\Lambda$ is {\em minus} one),
$\mu_S$ must take on a positive value to ensure a compensating
excess of kaons over antikaons,
leading to an approximate proportionality between $\mu_S$ and $\mu_B$.
This balancing of the strangeness in turn
leads to a small bias towards positive net charge
and $\mu_Q$ must therefore be correspondingly negative.
While the resulting value is also approximately proportional to $\mu_B$,
the absolute value is rather small 
and one might for most purposes simply take $\mu_Q$ to be zero.

In our spinodal scenario the value of $S_0$ is sampled at the plasma stage
and then canonically conserved though the hadronization process.
Each plasma blob has the volume $V_q$ (usually taken to be $50~\fm^3$)
and its strangeness content is determined by sampling
$\nu$ and $\bar\nu$ at the plasma temperature $T_q$.
which is either taken to be equal to $T_0$ ($=170\,\MeV$)
or to equal the hadronic freezeout temperature $T_h(\mu_B)$,
which should provide approximate upper and lower bounds.
At $\mu_B=0$, where $T_q=170~{\rm MeV}$, 
the width of the strangeness distribution
in such a blob is $\sigma_{S_0}=5.74$.
If we take $T_q=T_h$, the distribution will grow narrower
as $\mu_B$ is increased and at $\mu_B=400~{\rm MeV}$, 
where the temperature has decreased to 150~MeV, we have $\sigma_{S_0}=4.65$.

The blob then expands until the freezeout volume $V_h=\chi V_q$ is reached.
(Guided by rough approximations to the equation of state,
we usually employ an expansion factor of $\chi=3$, 
{\em i.e.}\ $V_h=150~\fm^3$.)
At this point, the blob has transformed itself into an assembly of hadrons 
whose abundances are assumed to be governed by the canonical distribution
(\ref{Pn}) associated with the particular value of $S_0$, and the 
modified chemical potentials $\mu_B'(S_0)$ and $\mu_Q'(S_0)$ 
(see Eq.\ (\ref{mu'})) that have been 
adjusted for each particular blob strangeness $S_0$ to ensure matching 
of the corresponding biased baryon and charge contents, as explained earlier.

\begin{figure}          
\includegraphics[angle=0,width=3.1in]{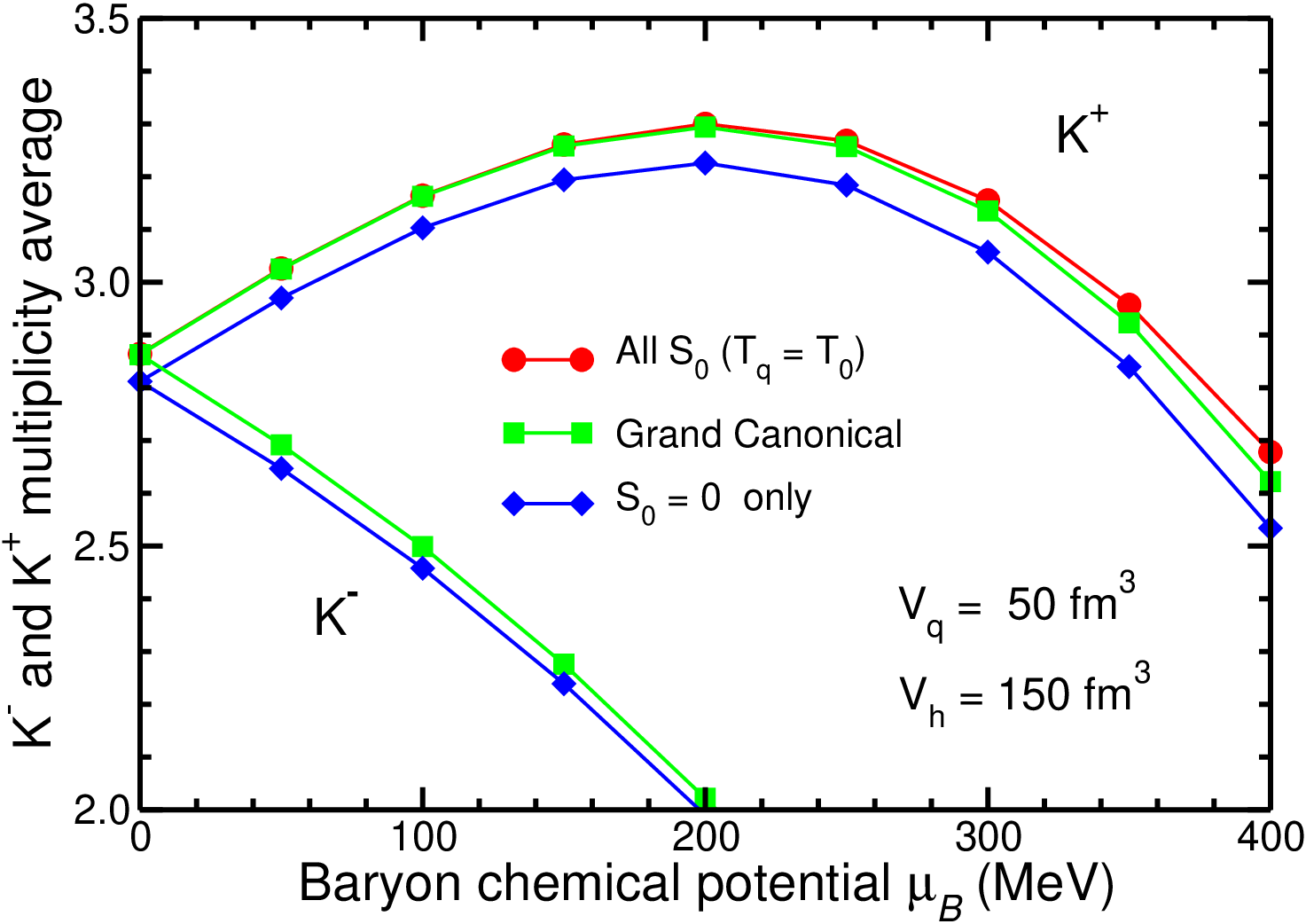}	
\caption{The average $K^\pm$ multiplicities
as functions of the baryon chemical potential $\mu_B$ in three scenarios:
1) the standard grand-canonical treatment where the hadronic freeze-out
occurs in the volume $V_h$ at the temperature $T_h(\mu_B)$,
2) our spinodal treatment in which the blob strangeness $S_0$
is sampled in the plasma volume $V_q$ at the temperature $T_q=T_0$
and then conserved through the hadronic freeze-out,
and 3) the restricted canonical treatment admitting only $S_0=0$.
}\label{f:Kave}
\end{figure}            

\subsection{Kaon multiplicity distributions}

We first discuss the results for the charged kaon multiplicity distributions.
Figure \ref{f:Kave} shows the resulting average $K^\pm$ multiplicities
for three different scenarios.
The first scenario is the usual global grand-canonical treatment (see above),
while the second is
our spinodal scenario in which the blob strangeness is determined
at the plasma stage and then kept fixed during the hadronization.
Both of these scenarios consider all possible values of $S_0$
but while the distribution of this quantity is determined 
at the hadronic freeze-out in the former,
it is determined already in the plasma in the latter.
In the third ``restricted canonical'' scenario 
the blob strangeness is always required to vanish, $S_0=0$.

The various scenarios lead to rather similar results.
The $K^-$ multiplicity decreases steadily as a result of the strangeness
balancing explained above while the $K^+$ yield initially increases,
for the same reason, but then, 
as the freeze-out temperature $T_h(\mu_B)$ begins to decrease noticeably, 
the overall hadron production decreases,
thus yielding a decreasing behavior of the $K^+$ curve.
Generally, the average multiplicities are affected very little
by the blob formation, relative to the grand-canonical treatment,
and the spinodal results are therefore omitted for $K^-$
while only the results for $T_q=T_0$ are shown for $K^+$.
These exhibit a progressively increasing enhancement
that amount to 2\% at $\mu_B=400\,\MeV$.
By contrast, for the restricted scenario, 
where only blobs having $S_0=0$ are admitted,
there is a reduction in the averages by a few per cent for all $\mu_B$.

\begin{figure}          
\includegraphics[angle=0,width=3.1in]{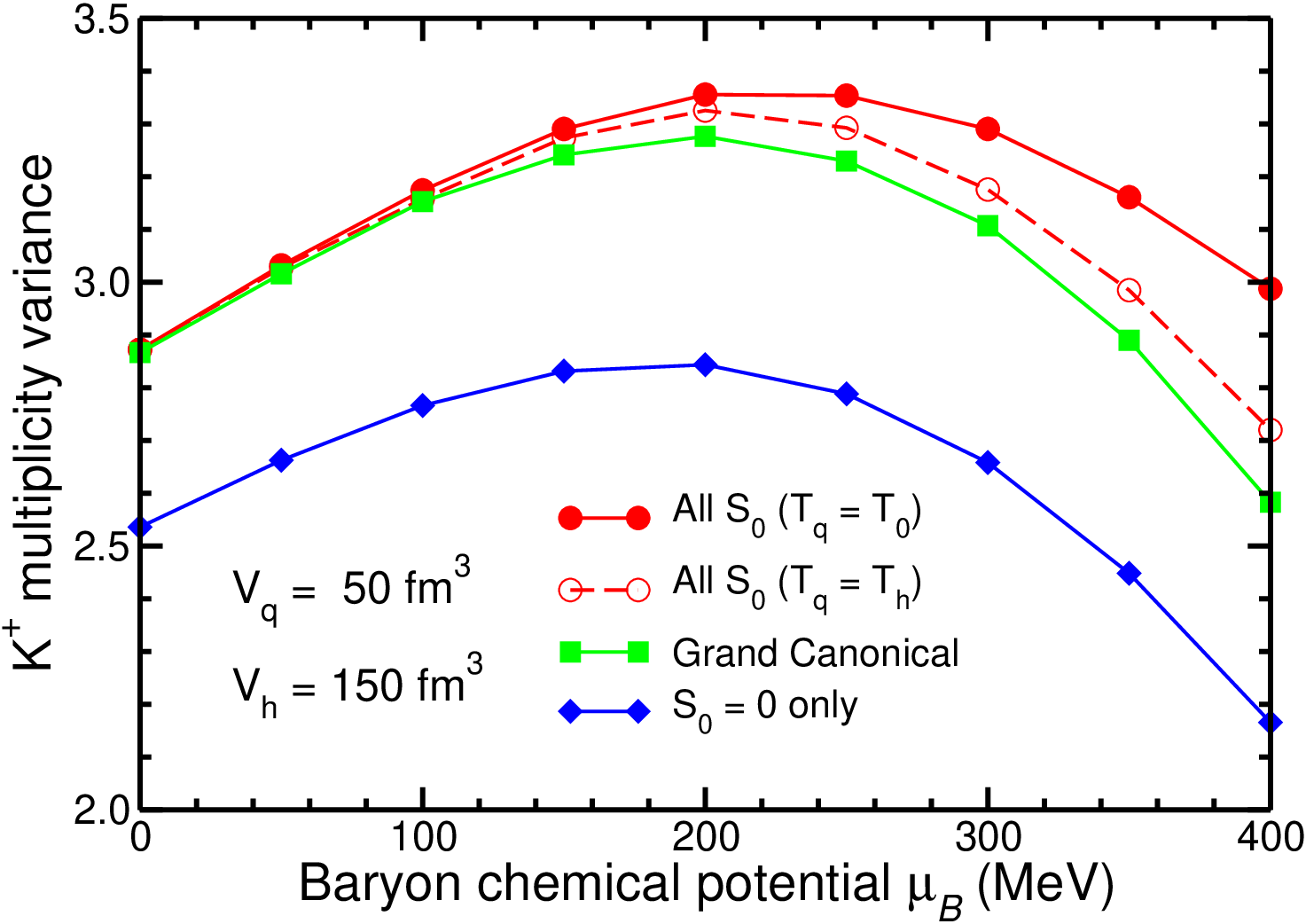}	
\caption{The variance in the $K^+$ multiplicity
as functions of the baryon chemical potential $\mu_B$
for the various scenarios considered in Fig.\ \ref{f:Kave},
using either $T_0$ or $T_h$ for the plasma temperature $T_q$
in the spinodal scenario.
}\label{f:Kvar}
\end{figure}            

The picture changes when the multiplicity fluctuations are considered,
as illustrated in Fig.\ \ref{f:Kvar}, where the corresponding
multiplicity variances are shown for the positive kaons.
While the overall behavior is qualitatively similar
to the behavior of the averages, 
there are several important differences.
First, in the restricted scenario (where only $S_0=0$ is included)
the suppression of the variance is significantly larger
than was the case for the average, amounting here to 8-10\%.
Furthermore, at the larger values of $\mu_B$,
where the net baryon density becomes significant,
the grand-canonical variance suffers more from the decreasing temperature
than the spinodal variance.
This important divergence is a result of the fact that
the larger average baryon number implies a correspondingly larger
baryon-number variance as well and therefore also a larger variance
in the strangeness (since strangeness in the plasma is carried exclusively 
by quarks and antiquarks which also carry baryon number).
Most importantly, there is a significant dependence
on the employed plasma temperature $T_q$.
In order to illustrate this effect,
we show results for two extreme values,
namely $T_q=T_0=170\,\MeV$ and $T_q=T_h(\mu_B)$,
and for these the difference amounts to ten per cent at $\mu_B=400~{\rm MeV}$.

\begin{figure}
\includegraphics[angle=0,width=3.1in]{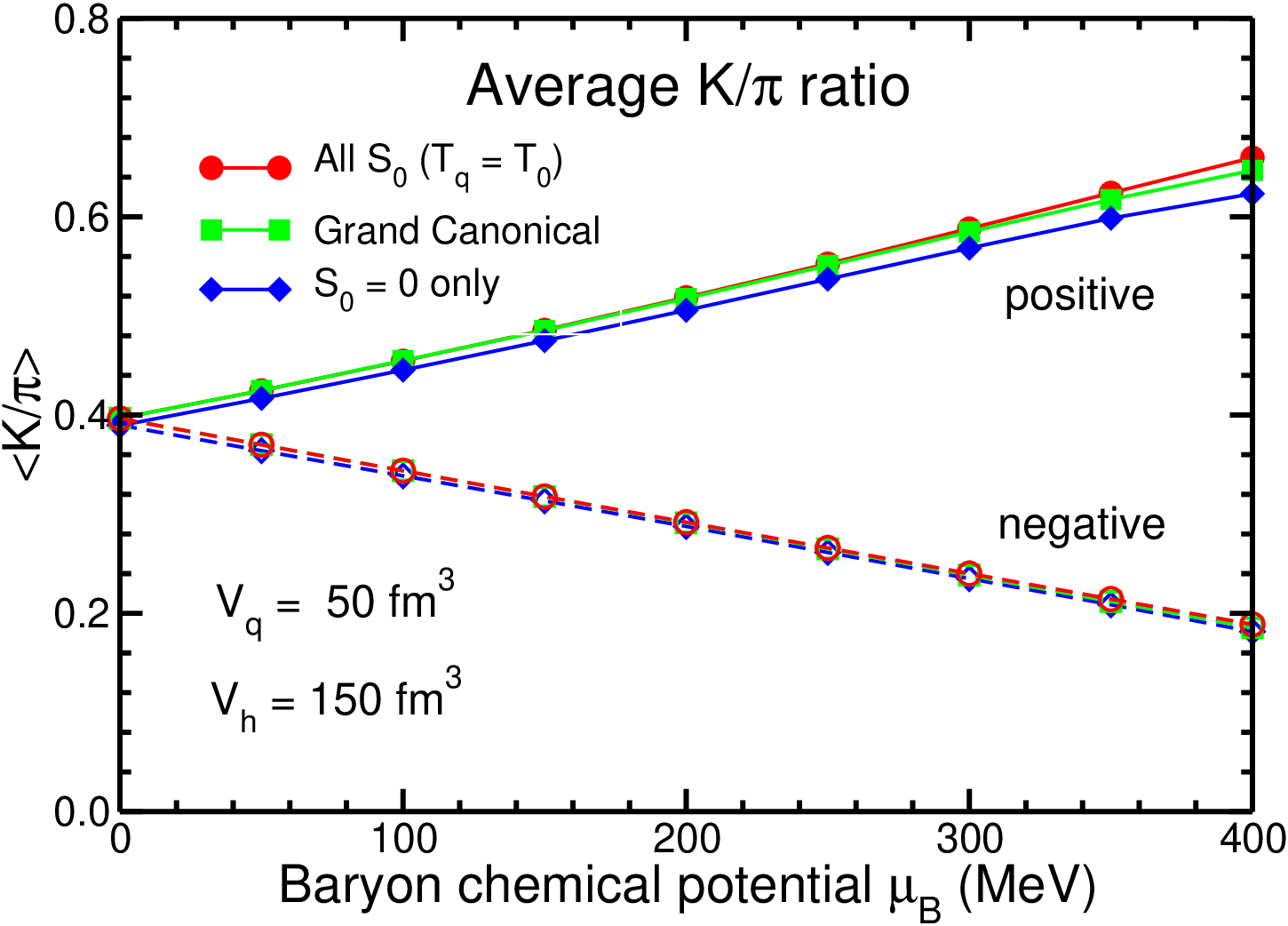}	
\caption{The average value of the $K/\pi$ ratio for 
either positive (increasing) or negative charges (decreasing)
for the three scenarios considered in Figs.\ \ref{f:Kave}-\ref{f:Kvar}.
}\label{f:aveK2pi}
\end{figure}            

\subsection{Multiplicity ratios}

From the experimental perspective, it is more convenient to consider
multiplicity {\em ratios}, and we therefore show in Fig.\ \ref{f:aveK2pi}
the average $K^+/\pi^+$ and $K^-/\pi^-$ ratios,
for the same three scenarios.
When $\mu_B$ is positive there is a preference for $K^+$ over  $K^-$
and hence $K^+/\pi^+$ will increase while $K^-/\pi^-$ decreases.
This behavior is practically linear since the suppression from the
decreasing temperature affects all hadrons species.
Since there is a (small) tendency for the $\pi$ and $K$ multiplicities
to vary in concert, the difference between the various scenarios is reduced.
In particular, there is hardly any difference to be seen
for the $K^-/\pi^-$ ratio.

\begin{figure}          
\includegraphics[angle=0,width=3.1in]{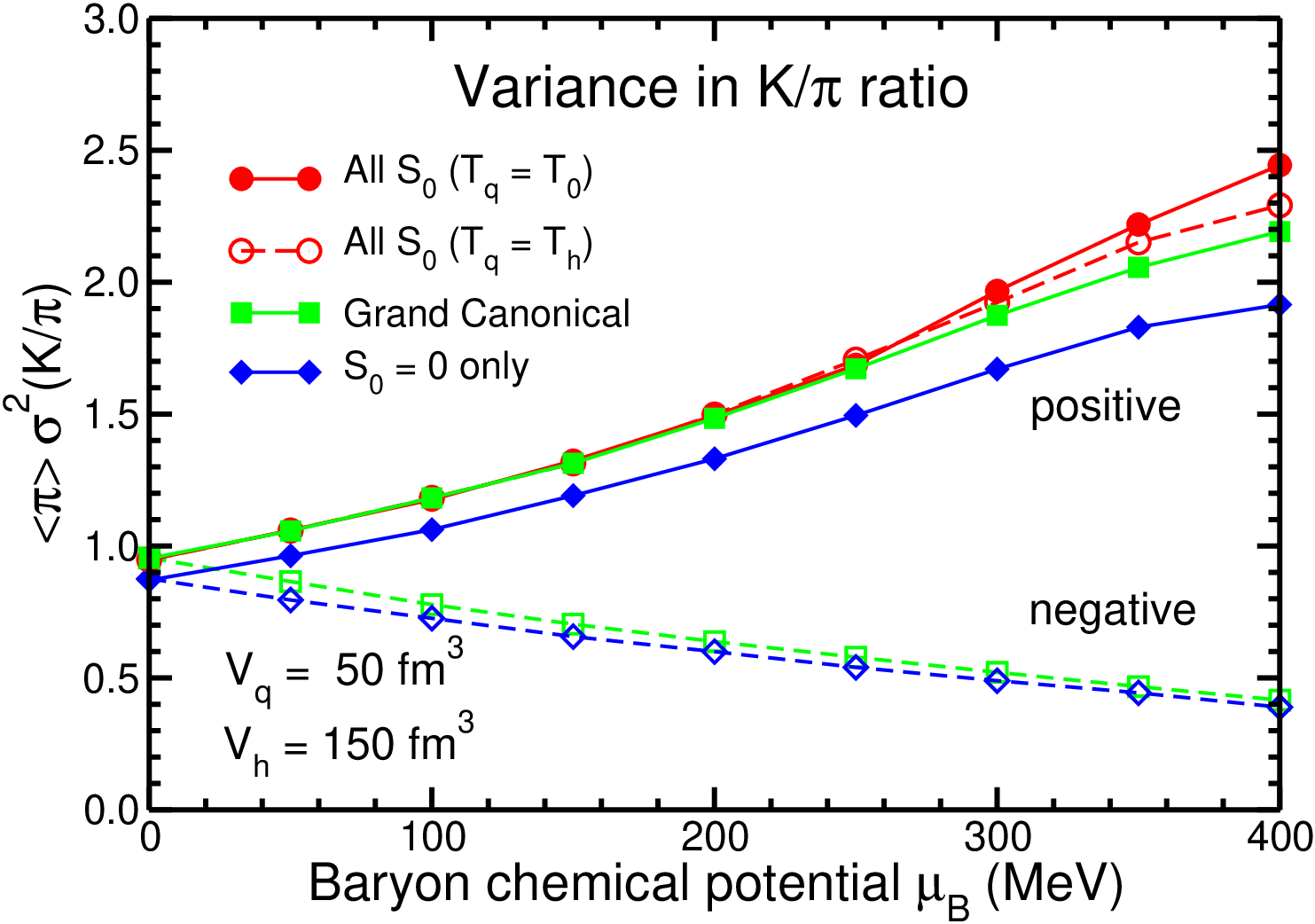}	
\caption{The normalized variance in the $K/\pi$ ratio for 
either positive (increasing) or negative charges (decreasing)
for the three scenarios considered in Figs.\ \ref{f:Kave}-\ref{f:Kvar}.
The variances have been multiplied by the corresponding 
average pion multiplicity in order to make the result scale invariant.
}\label{f:varK2pi}
\end{figure}            

We now turn to the corresponding variances.
Since the variance of the $K/\pi$ ratio
decreases in inverse proportion to the size of the system,
it is convenient to multiply by the mean pion multiplicity and thus
obtain a result that approaches a constant for large volumes,
$\langle\pi^\pm\rangle\sigma^2(K^\pm/\pi^\pm)$.
The resulting results for the positively charged hadrons
are shown in Fig.\ \ref{f:varK2pi}. 
They are qualitatively similar to those for $\langle K^\pm/\pi^\pm\rangle$
(Fig.\ \ref{f:aveK2pi}).
But although variances in the ratios are less sensitive to the specific 
scenario than the kaon variances themselves (Fig.\ \ref{f:Kvar}),
the differences are still clearly brought out.
Furthermore, as expected from Fig.\ \ref{f:Kvar},
an increase of the plasma temperature $T_q$ increases the resulting values.
The fluctuations in the $K^+/\pi^+$ ratio may thus offer a suitable
observable that is sensitive to a clumping-induced trapping of strangeness 
in the expanding matter prior to the hadronization.

\begin{figure}          
\includegraphics[angle=0,width=3.1in]{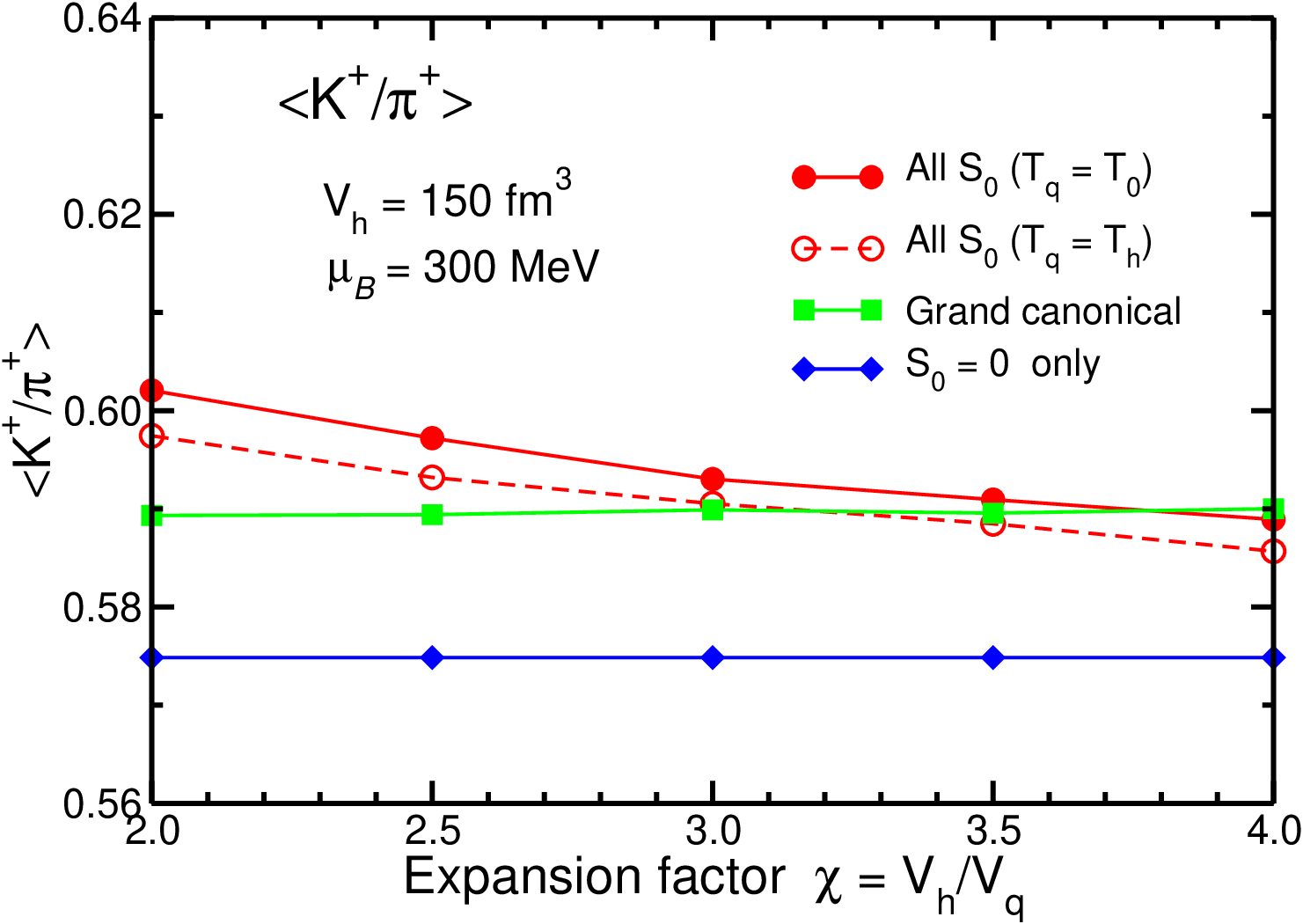}	
\includegraphics[angle=0,width=3.1in]{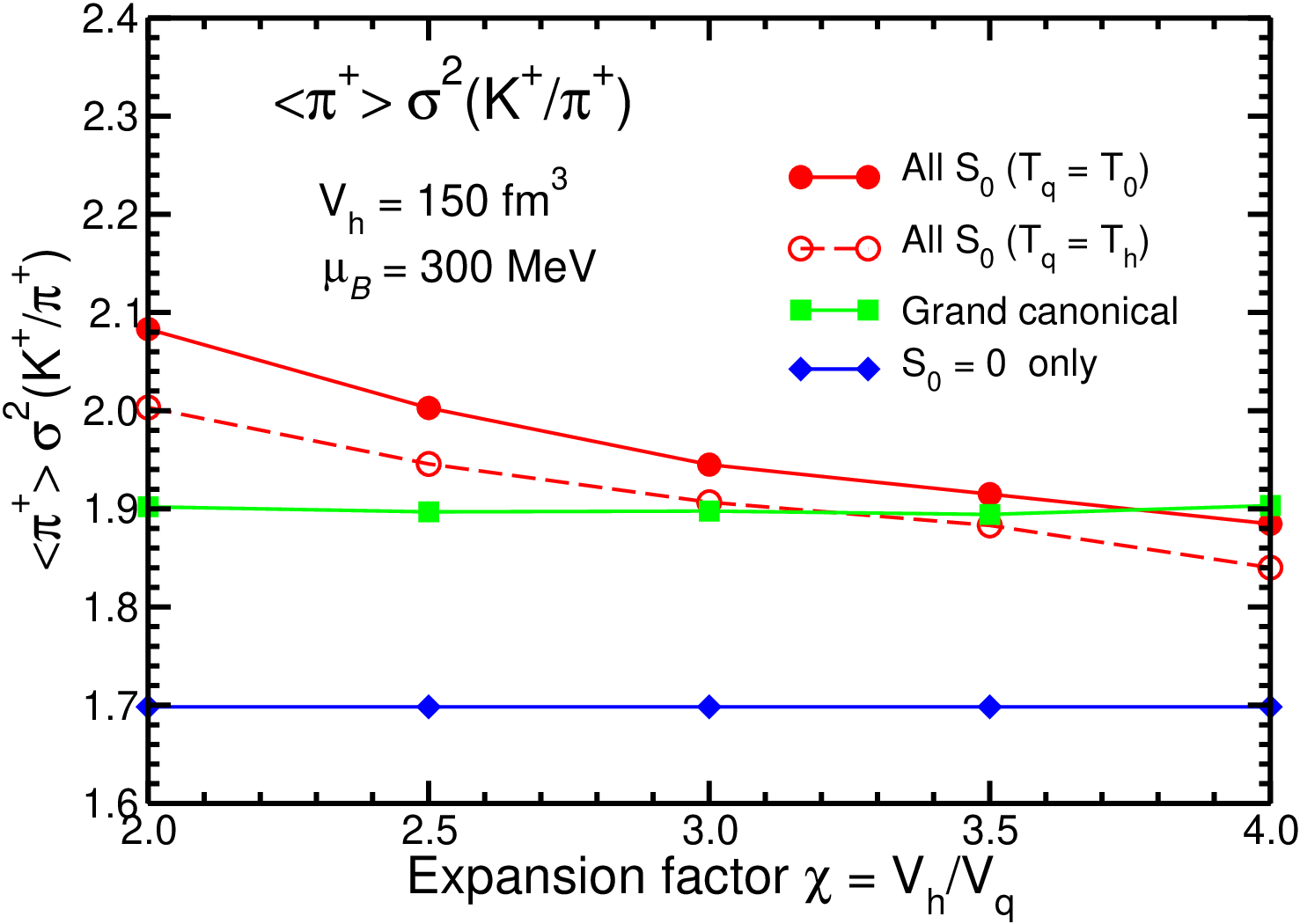}	
\caption{The average of the $K^+/\pi^+$ ratio (top)
and its variance multiplied by the mean $\pi^+$ multiplicity (bottom)
as obtained at $\mu_B=300~\MeV$
for the scenarios considered in Fig.\ \ref{f:varK2pi},
shown as a function of the volume expansion factor $\chi=V_h/V_q$.
}\label{f:chi}
\end{figure}		

\subsection{Dependence on the expansion factor}

The above results have been obtained for a given expansion factor 
$\chi\equiv V_h/V_q=3$.
This value should be taken only as a rough approximation to what might
actually happen and since the results are sensitive to this parameter
it is of interest to consider also other degrees of expansion.
This aspect is illustrated in Fig.\ \ref{f:chi},
where the average of the $K^+/\pi^+$ ratio
and its variance (normalized by $\langle\pi^+\rangle$)
are shown as a function of $\chi$ for $\mu_B=300~\MeV$.
Since we keep the hadronic freezeout volume equal to $V_h=150~{\rm fm}^3$
to facilitate the comparisons,
a larger expansion factor $\chi$ implies a smaller plasma volume $V_q$.
Since both averages and variances are proportional to volume, 
a smaller $V_q$ shrinks the distribution of the blob strangeness $S_0$
(so it has smaller mean and variance).
Consequently, 
the resulting values $\langle K^+/\pi^+\rangle$ 
and $\langle\pi^+\rangle\sigma^2(K^+/\pi^\pm)$
are decreasing functions of $\chi$.
However, this dependence is not dramatic:
a doubling of $\chi$ from 2 to 4 reduces the average ratio
by less than 2\% and the normalized variance by less than 10\%.

It is important to recognize that there are two opposing effects:
One is the basic fact there are more degrees of freedom
in the deconfined plasma phase than in the confined hadron gas, 
which enhances the fluctuations.
(At $T=T_0$ and $\mu_B=0$ we have $\zeta_s\approx0.33\,V_q$
while $\zeta_{S=\pm1}\approx0.081\,V_h$, 
so the effective degeneracy of the plasma
is approximately four times larger than that of the hadron gas,
which would then be compensated with an expansion factor of $\chi\approx4$,
as indeed borne out by the calculation.)
However, as $\chi$ is increased from unity,
this advantage is being steadily offset by the ever larger volume 
of the freeze-out configuration, $V_h=\chi V_q$.
The crossover happens to occur at $\chi\approx3.3$,
a value rather near our adopted estimate, $\chi=3$.
Should it turn out that there is less change in volume 
from the formation of the plasma blob to the hadronic freezeout,
then the relative effect of the strangeness trapping 
will be considerably larger.
For example, if $\chi$ were only ten per cent smaller,
the effect would be about twice as large.

\subsection{Effect of source mixing}

The above studies have considered only the hadrons 
resulting from a single blob which, presumably, 
populate a certain limited kinematical region centered
around the velocity of the original blob \cite{JR:HIP}.
However, even if a spinodal decomposition into 
kinematically separated blobs were to occur,
it would not be experimentally feasible to restrict the measurement
to include only those hadrons resulting from a single blob.
Rather, one should generally expect that a given detection acceptance 
will admit hadrons originating from more than one single blob.
Thus one needs to address the fact that a mixing of hadrons 
from different sources will degrade the strangeness-trapping signal.

\begin{figure}          
\includegraphics[angle=0,width=3.1in]{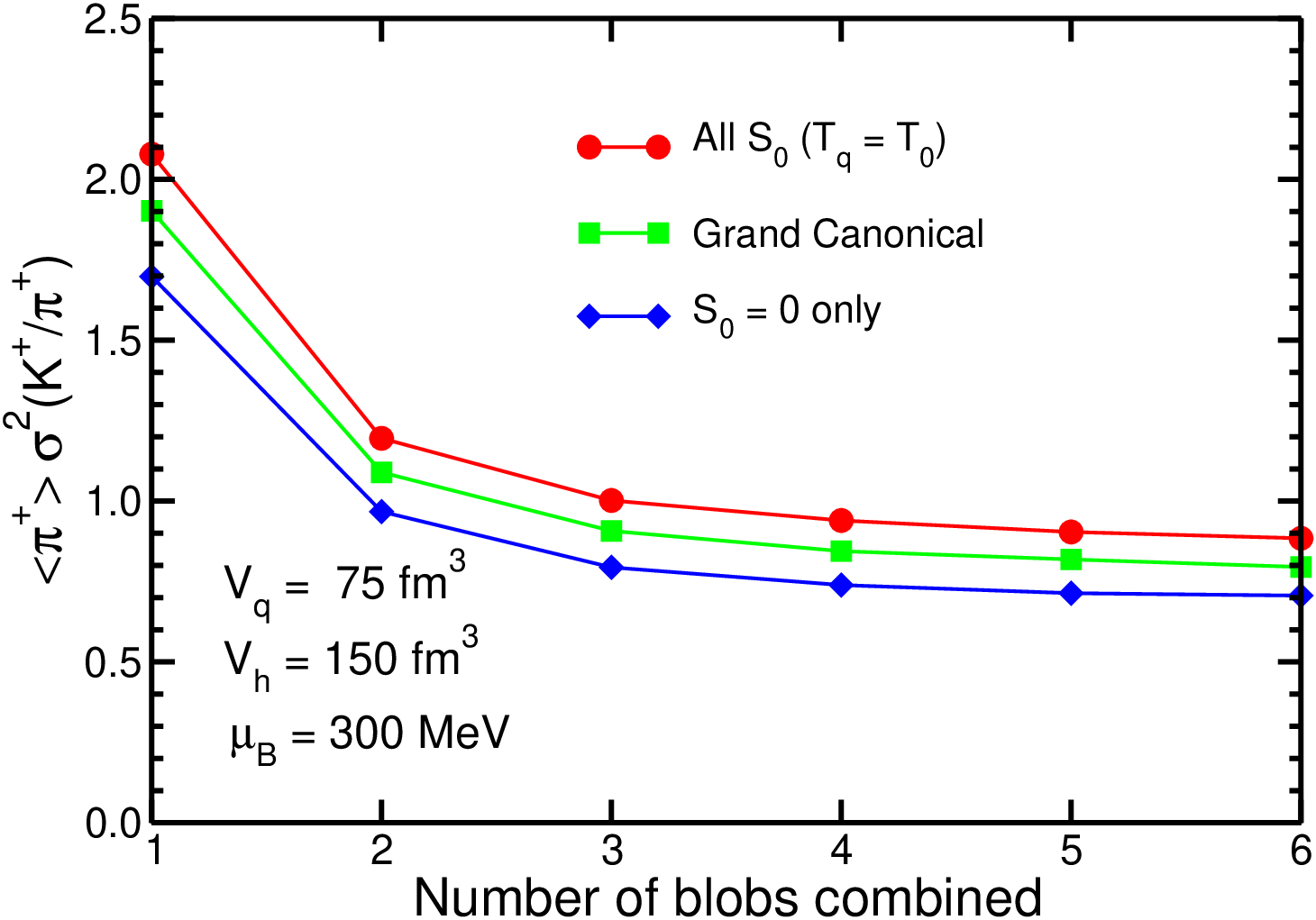}	
\caption{The effect of combining the hadrons from several separate 
plasma blobs before extracting the variance of the $K^+/\pi^+$ ratio,
for the three different treatments.
(For these results we used an expansion ratio of $\chi=2$.)
}\label{f:combine}
\end{figure}

In order to elucidate this practically important feature,
we calculate the production from several different blobs
and combine the resulting hadrons  
into a single ``event'' before performing the analysis.
The resulting variance in the $K/\pi$ ratio 
(multiplied by the mean $\pi^+$ multiplicity)
is shown in Fig.\ \ref{f:combine}
as a function of the number of blobs whose products have been combined.
While the value drops by about a factor of two
when going from a single blobs to two combined blobs,
it subsequently stabilizes and quickly approaches a constant
as ever more blobs are combined.
Furthermore, the relative increase when going from
the standard grand-canonical scenario one of our spinodal scenarios
remains nearly unaffected by the numbers of blobs combined in the analysis.
This result indicates that the suggested signal of the strangeness trapping
is robust against the inevitable source mixing and, consequently,
it may in fact be practically observable.

\subsection{Global strangeness conservation}
\label{sec:canonical}

We finally analyze in some detail the role played by the overall
conservation of strangeness in each event.

In the above studies, we have treated the strangeness in a given plasma blob
as a grand-canonical variable,
which is well justified when the blob volume forms only a small part
of the total system.
In order to investigate the quality of this approximation,
we consider the effect of constraining the combined strangeness 
of $N$ individual blobs to zero.
The corresponding canonical partition function for the combined system is then
\beq
\Z_0^{(1\cdots N)}\ =\
\prod_{i=1}^N\left[\sum_{S_i} \Z_{S_i}^{(i)}\right] 
\delta_{S_1+\ldots+S_N,0}\ ,
\eeq
where $\Z_{S_i}^{(i)}$ is the canonical partition function for blob $i$
having the strangeness $S_i$.
For simplicity, we shall assume that all $N$ blobs are similar
(as we have done above).
First, to obtain the correlated distribution 
of the blob strangenesses $(S_1,\dots,S_N)$,
we consider the problem at the plasma level where $\Z_0=I_0(2N\zeta_s)$.
We then find 
\beq
P_0(S_1,\ldots,S_N)\ =\ {\delta_{S_1+\ldots+S_N,0}\over I_0(2N\zeta_s)}
\prod_{i=1}^N\left[I_{S_i}(2\zeta_s)\right]\ .
\eeq
as described in more detail in Appendix \ref{global}.

In order to bring out the effect of the global strangeness constraint,
we consider the multiplicity of a particular strange hadron $\kappa$
resulting from the combined observation of some of the blobs,
say those labeled $1,\dots,N'$,
{\em i.e.}\ $n_\kappa=n_\kappa^{(1)}+\ldots+n_\kappa^{(N')}$,
where $n_\kappa^{(i)}$ is the multiplicity contributed by the blob $i$.
If all the blobs are similar, the average multiplicity is of the form
\beq\label{ave}
\langle n_\kappa\rangle = N' n_\kappa^{\rm ave}(N)\ ,\
\eeq
where $n_\kappa^{\rm ave}(N)$ is the average multiplicity arising
from any single blob.
Furthermore, the variance in the total multiplicity $n_\kappa$
has the following form,
\beq\label{var}
\sigma_\kappa^2\ 
=\ N'[\sigma_\kappa^{\rm var}(N) + 
{N'-1 \over N-1} \sigma_\kappa^{\rm cov}(N)]\ ,
\eeq
where $\sigma_\kappa^{\rm var}(N)$ is the variance in multiplicity 
from any single source and $\sigma_\kappa^{\rm cov}(N)$ is the covariance 
between the multiplicity from any one source
and the combined multiplicity from the all the other $N-1$ sources.
The overall restriction of the total strangeness to zero
reduces the individual multiplicities $\{n_\kappa^{(i)}\}$,
leading to smaller values of 
both $n_\kappa^{\rm ave}$ and $\sigma_\kappa^{\rm var}$.
Furthermore, a higher-than-average strangeness in one blob
introduces a bias towards lower-than-average values in the others,
in that particular ``event'',
thus producing an anticorrelation among the individual strangeness values
$\{S_n\}$.
This in turn leads to negative covariances, $\sigma_\kappa^{\rm cov}<0$.

\begin{figure}          
\includegraphics[angle=0,width=3.1in]{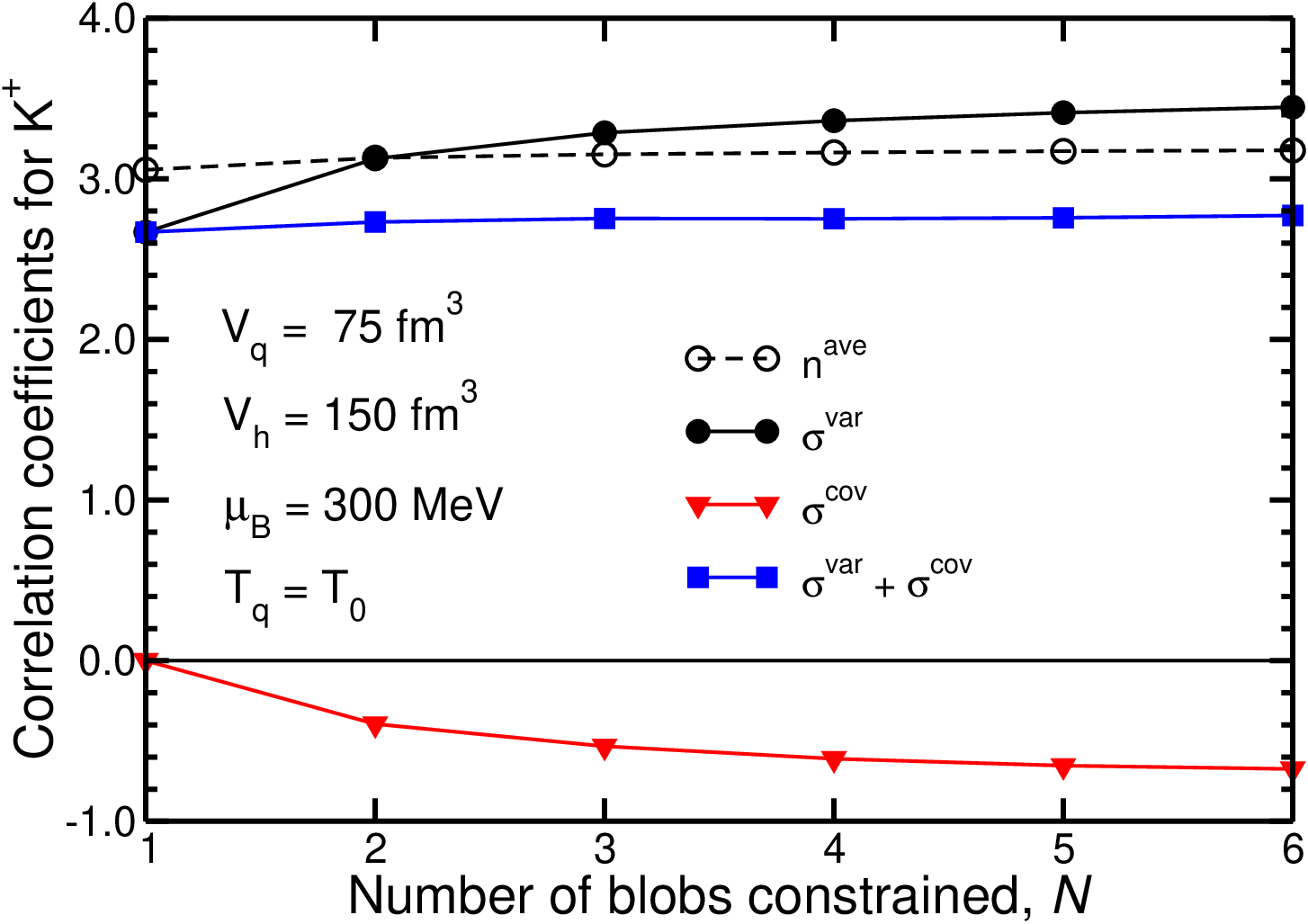}	
\caption{The dependence of the coefficients 
$n^{\rm ave}$, $\sigma^{\rm var}$, $\sigma^{\rm cov}$
on the total number of blobs that are subject to the 
global strangeness constraint, $S_{\rm tot}\!=\!0$,
for $K^+$ in a scenario having
$\mu_B=300~\MeV$, $T_q=T_0$, and $\chi=2$,
as in Fig.\ \ref{f:combine}.
}\label{f:global}
\end{figure}            

To illustrate these effects, we consider the production of positive kaons 
and show in Fig.\ \ref{f:global} the dependence of the coefficients
$n^{\rm ave}$, $\sigma^{\rm var}$, and $\sigma^{\rm cov}$ on $N$,
the total number of blobs that are subject to the global constraint.
(We may here employ either analytical recursion relations 
or statistical simulation, as discussed in Appendix \ref{global}.)
For $N\!=\!1$ the strangeness of each blob must vanish and we obtain
the restricted canonical scenario considered earlier.
In the opposite extreme, $N\to\infty$, 
the global constraint becomes ineffective and
the standard grand-canonical scenario is recovered.
To a good approximation, the deviations of the coefficients
from their grand-canonical values are inversely proportional to $N$,
for example,
\beq
\sigma_\kappa^{\rm var}(N)\ \approx\ \sigma_\kappa^{\rm var}(\infty)\ +\
\frac{1}{N}[\sigma_\kappa^{\rm var}(1)-\sigma_\kappa^{\rm var}(\infty)]\ .
\eeq
One may then readily judge the effect for a given $N$
on the basis of the change in going from $N=\infty$ to $N\!=\!1$.
For the average number of $K^+$ emitted by each blob, $n^{\rm ave}$,
this drop is only about 4.5\% (2.5\%),
while it amounts to 25\% (16\%) for the corresponding variance, 
$\sigma^{\rm var}$ (the numbers in paranthesis refer to $\chi$=3).
This behavior may be contrasted with the approximate constancy of 
$\sigma^{\rm var}+\sigma^{\rm cov}$,
the variance of the total $K^+$ multiplicity divided by $N$,
which decreases by only 4\% (2\%) when going from $N\gg1$ to $N\!=\!1$.
These results illustrate the fact that the variance in the total multiplicity
is always smaller than the corresponding mean when the system is subject to
a canonical constraint.

\begin{figure}          
\includegraphics[angle=0,width=3.1in]{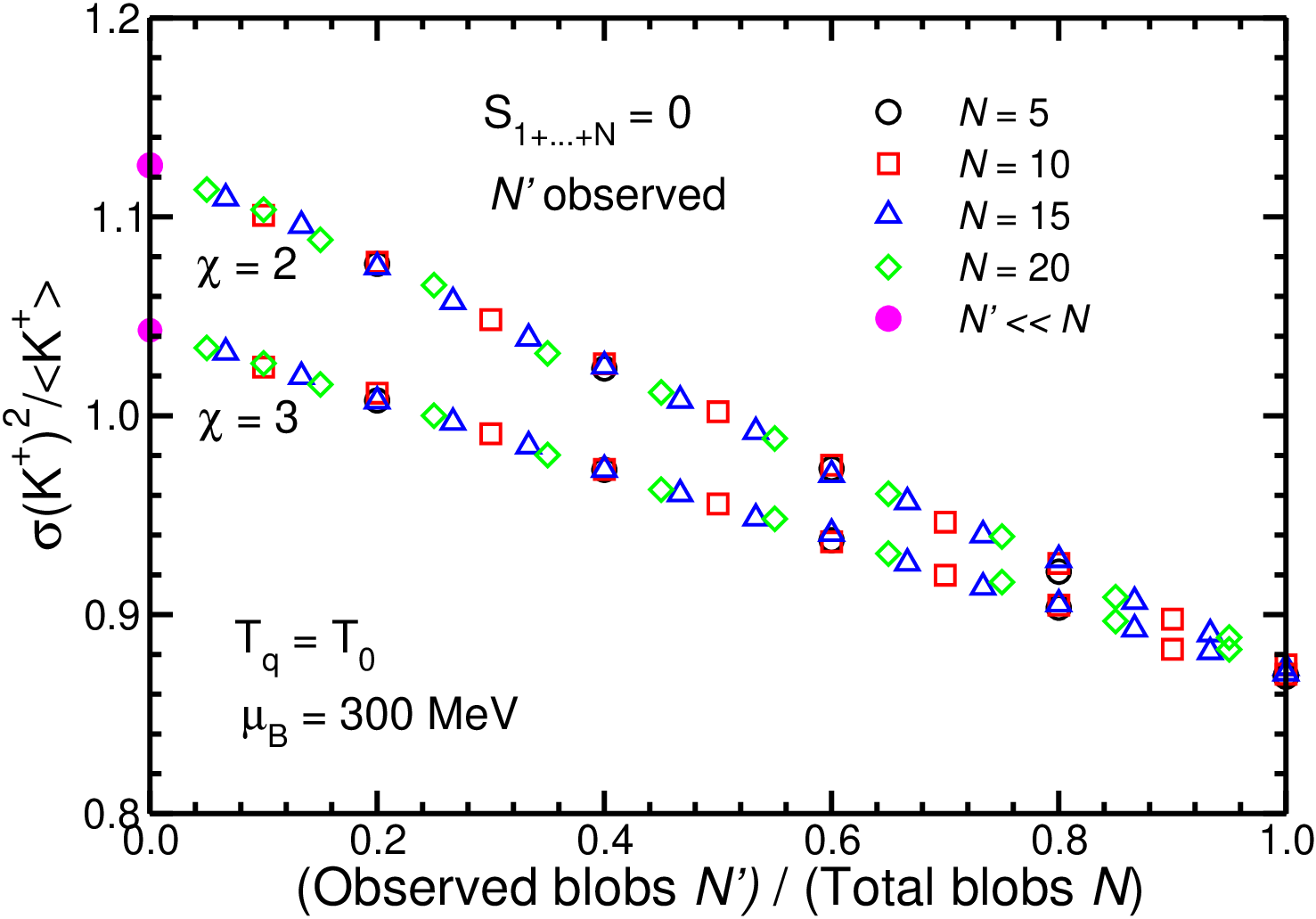}	
\caption{The ratio of the variance to the mean
for the $K^+$ multiplicity resulting from a combined system of $N$ blobs
of which $N'$ are being observed, plotted as a function of $N'/N$.
These results were obtained for $\mu_B=300~\MeV$ and $T_q=T_0$
with either $\chi=2$ (upper curve) or $\chi=3$ (lower curve).
}\label{f:canonical}
\end{figure}            

Figure \ref{f:canonical} illustrates the dependence
of the canonical effect on the ratio $N'/N$
by displaying the ratio of the variance to the mean
for the $K^+$ multiplicity resulting from a combined system of $N$ blobs
of which $N'$ are being observed.
For any given value of the expansion ratio $\chi=V_h/V_q$,
which governs the effective statistical weight of the plasma 
relative to that of the hadron gas,
the change from the limit where only a small fraction of the combined system 
is being observed to the situation when all the kaons are collected
is seen to follow a universal curve that is linear in $N'/N$, 
to a very good aproximation.

\begin{figure}          
\includegraphics[angle=0,width=3.1in]{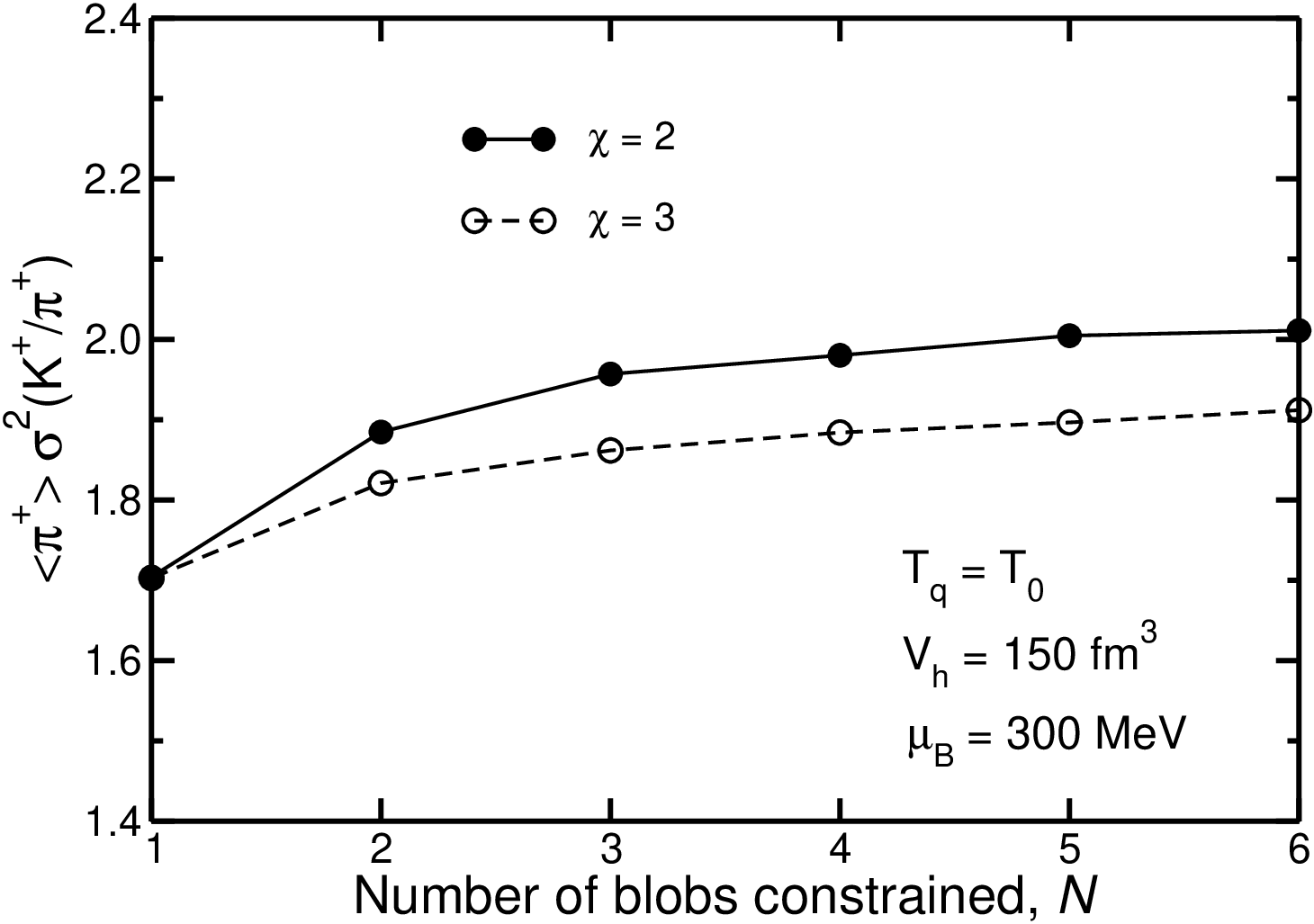}	
\caption{The variance of the $K^+/\pi^+$ ratio for a single blob,
multiplied by the mean $\pi^+$ multiplicity,
when the combined net strangeness of $N$ blobs is required to vanish.
These results were obtained for $\mu_B=300~\MeV$ and $T_q=T_0$
with either $\chi=2$ (upper curve) or $\chi=3$ (lower curve),
as in Fig.\ \ref{f:canonical}.
}\label{f:constrain}
\end{figure}            

The corresponding effect on the fluctuations of the $K/\pi$ ratio
is illustrated in Fig.\ \ref{f:constrain}.
It shows the value of $\langle\pi^+\rangle\sigma^2(K^+/\pi^+)$
obtained for the hadrons emitted from a single blob 
when $N$ such blobs satisfy a combined canoncial constraint 
on their total strangeness contents, $S_{\rm tot}=0$.
The canonical constraint is obviously most effective 
in the extreme case of a single blob, $N\!=\!1$,
where it is identical to the restricted scenario
in which only $S_0=0$ is allowed for each individual blob
(bottom curve in Fig.\ \ref{f:combine}).
As more blobs share the burden of the constraint,
its effect rapidly diminishes 
and the results approach those of our standard spinodal scenario,
where the strangeness of each individual plasma blob is determined
grand canonically (top curve in Fig.\ \ref{f:combine});
this approach occurs approximately as $\sim1/N$.
Thus already at $N\approx5$ the effect is hardly significant.
Consequently, if a given blob represents less than 20\% of the total system,
the vanishing of the overall strangeness should have no noticeable effect
on the observables considered here.

\section{Concluding discussion}

Most observables considered so far in high-energy collision experiments
behave rather smoothly as functions of the control parameters, 
such as bombarding energy and centrality.
However, preliminary data analysis of fixed-target experiments at the SPS
by the NA49 collaboration \cite{NA49_kp} have revealed two 
intriguing exceptions. 
First, there appears to be a significant enhancement of the $K/\pi$ 
ratio at beam energies of $20-30~A\GeV$.
Second, in the same energy range, the fluctuation of this ratio
(expressed relative to mixed events) is strongly enhanced \cite{becattini}.
Both enhancements are localized within about $10~A\GeV$ in beam energy. 
Though otherwise rather successful in describing the hadron yields,
statistical models cannot reproduce the observed energy dependence
of these enhancements \cite{becattini,oeschler}. 
Nor does a hadron gas and its resonance-induced correlations account for
the strong enhancement in the ratio fluctuations \cite{jeon1,jeon2}. 
One might then wonder whether these strong fluctuations 
might result from the enhanced fluctuations 
associated with a second-order phase transition \cite{Stephanov:1998dy}.
However, if this were the case, 
one would expect particularly strong fluctuations in the pions. 
However, while the fluctuations of the $K/\pi$ ratio are 
observed to be enhanced, those of the $p/\pi$ ratio 
follow the expectations from transport models \cite{Roland}. 

If the statistical models \cite{PBM} 
indeed provide reasonable estimates for the relation between beam energy 
and the thermodynamic parameters characterizing the chemical freeze-out,
{\em i.e.}\ the temperature and the chemical potentials,
then the anomalous behavior reported by NA49
would be consistent with a first order-phase transition 
having occurred prior to the chemical freeze-out, 
at a somewhat higher temperature and about the same chemical potential.
In order to evaluate the plausibility of such a speculation,
we have in this paper studied the effects on yields and fluctuations
if the bulk of the system indeed undergoes a spinodal-like break-up
as it hadronizes.
The key physical effect is then that the amount of strangeness
residing within the plasma that forms a given blob remains effectively trapped,
thus imposing a canoncial strangeness constraint 
on its conversion into hadrons.

Considering idealized scenarios in which an equilibrated plasma forms 
a number of separate blobs that subsequently expand and hadronize
independently, we have developed the relevant statistical tools.
These results may be of general applicability.
With a view towards the specific observations at the SPS,
we have studied the production of pions and kaons in such a scenario
relative to the standard picture in which statistical equilibrium
is maintained through freeze-out.
While the average hadron yields are essentially unaffected by the breakup,
the fluctuations in the strange-particle multiplicities are significantly 
enhanced, especially at larger values of the chemical potential.
From the experimental perspective, 
the yield {\em ratios} are particularly interesting
and we have especially studied the $K/\pi$ ratio.
Depending on the degree of expansion during the stage
of strangeness trapping, the fluctuations in the $K^+/\pi^+$ ratio
can be enhanced significantly (of the order of 10\%)
relative to the standard scenario where global equilibrium is maintained.

Thus, the present (rather idealized) studies suggest that a spinodal
decomposition might indeed lead to enhancements of the magnitude
observed by NA49.
However, before such a conclusion could be made with confidence, 
further studies would be required.
In particular, both strong resonance decays and weak decays
must be taken into account.
Moreover, the enhancement of the $K/\pi$ fluctuations
should be correlated with other expected effects, 
such as $N$-body kinematic correlations \cite{JR:HIP}.
Of particular interest are more precise calculations 
of the equation of state in the relevant baryon-rich environments,
to better ascertain the location of the expected phase coexistence region,
as well as refined calculations of the collision dynamics
to determine whether the spinodal region is in fact likely to be encountered
and, if so, to what degree the phase decomposition may actually develop.
We hope that the present findings will provide added incentive 
for these challenging undertakings.

\appendix
\section{Canonical fermions}
\label{fermions}

We discuss here the canonical treatment of fermions,
in which the total number of quanta present
is kept fixed to a given value $N_0$ (the number of ``particles).
The appropriate partition function is then
\beqar
\Z_{N_0}(\beta) &=& 
\prod_i\left[\sum_{n_i=0}^1\rme^{-\beta n_i\varepsilon_i}\right]
\delta(\sum_i n_i-N_0)\\ &=& \nonumber
\sum_{\{n_i\}_{N_0}} \rme^{-\beta E\{n_i\}}\ ,\
E\{n_i\} = \sum_i n_i\varepsilon_i\ .
\eeqar
Here $n_i$ is the number of quanta present in the particular 
``single-particle'' state $i$ of energy $\varepsilon_i$.
Furthermore, $\{n_i\}_{N_0}$ denotes the subset of all configurations 
$\{n_i\}$ whose total particle number is constrained
 to be equal to the specified value $N_0$,
$N\{n_i\}\equiv\sum_i n_i\doteq N_0$.
The fermionic restriction on $n$ to be at most one
complicates the evaluation of the partition function.
However, it is possible to derive the following recursion relation,
\beq
\Z_N^F(\beta) = {1\over N}\sum_{n=1}^N 
(-)^{n-1} \Z_{N-n}^F(\beta)\, \Z_1^F(n\beta)\ ,
\eeq
with $\Z_0^F(\beta)=1$.
$N$ can be readily sampled numerically on the basis of
$P(N)=\Z_N/\Z$, where $\Z^F = \sum_{N\geq0}\Z_N^F$ 
is the grand-canonical partition function.

One way to derive the above recursion relation 
reorganizes the basic expression for $Z_N$,
\beqar
Z_N &=& \sum_{i_1<i_2<\cdots<i_N}\rme^{a_{i_1}+a_{i_2}+\cdots+a_{i_N}}\\
&=& {1\over N!}  \sum_{i_1\neq i_2\neq\cdots\neq i_N}
\rme^{a_{i_1}+a_{i_2}+\cdots+a_{i_N}}\ ,
\eeqar
where $a_i\equiv-\beta\epsilon_i$ 
and no indices may be equal in the second sum. 
With $\zeta_n\equiv\Z_1(n\beta)$, the first few terms are
\beqar\nonumber
\Z_2\!\!&=&\!\! {1\over2!} \sum_{i\neq j}\rme^{a_{i}+a_{j}} =
\half \left[\sum_{i j}\rme^{a_{i}+a_{j}} 
-\sum_i \rme^{2a_i} \right]\\
&=& \half \left[\Z_1(\beta)^2-\Z_1(2\beta)\right]
 = \half \left[\Z_1\zeta_1-\Z_0\zeta_2\right]\ ,\\ \nonumber
\Z_3\!\!&=&\!\! {1\over3!}
\sum_{i\neq j\neq k}\rme^{a_{i}+a_{j}+a_{k}}
= {1\over3!}
\left[\sum_{ijk}\rme^{a_{i}+a_{j}+a_{k}} -\dots\ \right]\\
&=& \third[\Z_2\zeta_1-\Z_1\zeta_2+\Z_0\zeta_3]\ ,
\eeqar
The first term is the classical limit,
$\Z_N=\Z_1^N/N!+\cdots$.

Another approach starts from the expansion of $\ln\Z$,
\beqar\label{lnZ}\nonumber
\ln\Z &=& \sum_i\ln[1+\rme^{-\beta\epsilon_i}]
= \sum_i \sum_{n\geq0} (-)^{n-1}{1\over n}\, \rme^{-n\beta\epsilon_i}\\
&=& \sum_{n\geq0} (-)^{n-1}{1\over n} \zeta_n
= \zeta_1-\half\zeta_2+\third\zeta_3+\cdots\ ,
\eeqar
then exponentiates $\ln\Z=[\zeta_1-\half\zeta_2+\third\zeta_3+\cdots]$,
\beqar
&\Z& =\ \rme^{[\,\cdot\,]}\ =\ \sum_{N\geq0} {1\over N!} [\,\cdot\,]^N\\
\nonumber
&=&\! 1 + [\,\cdot\,]\left(1+\half[\,\cdot\,]\left((1+\third[\,\cdot\,]
\left((1+\quart[\,\cdot\,]\left((1+\dots\right)\right)\right)\right) ,
\eeqar
and finally reorganizes the series in powers of $\rme^{\beta\epsilon}$,
\beqar
\Z &=& \sum_{N\geq0}\Z_N\ =\
1\ +\ \zeta_1\ +\ \underbrace{\half[\zeta_1^2-\zeta_2]}_{\Z_2}\\ \nonumber
&~& + \third[\underbrace{\half(\zeta_1^2-\zeta_2)}_{\Z_2}\zeta_1
-\zeta_1\zeta_2+\zeta_3]\ +\ \dots\  .
\eeqar

\paragraph{Degeneracy and volume.}
The above derivations have been made for non-generate systems, $g=1$.
In the general case, when $g>1$, the situation is more complicated
since a given energy level may accommodate up to $g$ quanta.
There is then no simple relation between $\Z[g=1]$ and $\Z'\equiv\Z[g>1]$.  
However, using the above expansion (\ref{lnZ})
of $\ln\Z$ in terms of $\zeta_n$ we find
\beq
\ln\Z'\ =\ g\sum_i\ln[1+\rme^{-\beta\epsilon_i}]\ =\ \dots\
=\ \zeta_1'-\half\zeta_2'+\third\zeta_3'+\cdots\ ,
\eeq
where $\zeta_n'\equiv g\zeta_n=g\Z_1(n\beta)$.
The partition function may thus be obtained by replacing 
$\zeta_n$ by $\zeta_n'=g\zeta_n$ in the above procedure,
\beq
\Z'\ =\ \sum_N \Z_N'\ ,\,\ \Z_N'\ = \Z_N\{\zeta_n\to\zeta_n'=g\zeta_n\}\ .
\eeq
Consequently, the recursion relation takes the form,
\beq
\Z_N' = {1\over N}\sum_{n=1}^N 
(-)^{n-1} \Z_{N-n}'\, \zeta_n'\ ,\ \Z_0=1\ , 
\eeq
with $\zeta_n' \equiv g\zeta_n = g\sum_i\rme^{-n\beta\epsilon_i}$,
and $\Z_N'$ becomes an $N^{\rm th}$ order polynomial in the degeneracy $g$,
\beqar
&~& \Z_0'=1\ ,\ \Z_1'=g\zeta_1\ ,\
\Z_2'=\half g^2\zeta_1^2-\half g\zeta_2\ ,\\
&~& \Z_3'=\sixth g^3\zeta_1^3-\half g^2\zeta_1\zeta_2
	+\third g\zeta_3\ ,\cdots\ .
\eeqar
We note that the dependence on the {\em volume} $V$ 
is similar to the dependence on the degeneracy $g$, 
since both enter as overall factors in the
elementary partition functions, $\zeta_n\sim gV$.

\paragraph{Multiplicity distribution.}

The mean multiplicity is enhanced/reduced for bosons/fermions,
\beqar\nonumber
\langle N\rangle &=& {1\over\Z}\sum_N N\Z_N =
{g\over2\pi^2}{VT^3\over\hbar^3c^3}\sum_{n>0}(\pm)^{n-1}{1\over n^3}
\tilde{K}_2({nm\over T})\\
&=& \sum_{n>0}(\pm)^{n-1}\zeta_n =\zeta_1\pm\zeta_2+\cdots\ ,
\eeqar
and so is the corresponding multiplicity variance,
\beqar
\sigma_N^2 &=& 
{g\over2\pi^2}{VT^3\over\hbar^3c^3}\sum_{n>0}(\pm)^{n-1}{1\over n^2}
\tilde{K}_2({nm\over T})\\ \nonumber
&=& \sum_{n>0}(\pm)^{n-1} n\zeta_n =\zeta_1\pm2\zeta_2+3\zeta_3\pm\cdots\ ,
\eeqar
and both are strictly proportional to both the volume $V$
and the degeneracy $g$, since $\zeta_n\sim gV$,
as just noted above.

\paragraph{Chemical potentials.}
The above results apply in the absence of a chemical potential.
The presence of a chemical potential effectively replaces 
the energy $\epsilon_i$ by a shifted value $\epsilon_i-\mu$
and consequently all the manipulations can be carried through as above.
So, with
\beq
\zeta_n(\alpha,\beta) \equiv \ZERO{\zeta}_n(\beta)\,\rme^{-\alpha n}\ ,\
\ZERO{\zeta}_n(\beta) \equiv \Z_1(n\beta) = \sum_i\rme^{-n\beta\epsilon_i}\ ,
\eeq
we see that all the terms in $\Z_N$ contain the same power of
the fugacity $\rme^{-\alpha}$ and we find
\beq
\Z_N(\alpha,\beta)\ =\ 
\ZERO{\Z}_N\{\ZERO{\zeta}(n\beta)\}\,\rme^{-\alpha N}\ .
\eeq
However, the multiplicity moments have a complicated $\alpha$ dependence,
\beqar
\langle N\rangle &=& \ZERO{\zeta}_1\,\rme^{-\alpha}
\pm\ZERO{\zeta}_2\,\rme^{-2\alpha}
+\ZERO{\zeta}_3\,\rme^{-3\alpha}\pm\dots\ ,\\
\sigma_N^2 &=& \ZERO{\zeta}_1\,\rme^{-\alpha}
\pm2\ZERO{\zeta}_2\,\rme^{-2\alpha}
+3\ZERO{\zeta}_3\,\rme^{-3\alpha}\pm\cdots\ .\,\,\
\eeqar
Furthermore, the corresponding expressions for the associated antiparticle are
\beqar
\langle\bar{N}\rangle &=& \ZERO{\zeta}_1\,\rme^{+\alpha}
\pm\ZERO{\zeta}_2\,\rme^{+2\alpha}
+\ZERO{\zeta}_3\,\rme^{+3\alpha}\pm\dots\ ,\\
\sigma_{\bar N}^2 &=& \ZERO{\zeta}_1\,\rme^{+\alpha}
\pm2\ZERO{\zeta}_2\,\rme^{+2\alpha}
+3\ZERO{\zeta}_3\,\rme^{+3\alpha}\pm\cdots\ .\,\,\
\eeqar

In the case of a plasma blob, which has no strangeness bias,
the distributions of $N$ and $\bar N$ must be identical.
Consequently, in that situation, we must have $\alpha=0$.
Thus, if the $s$ and $\bar s$ quarks are embedded in an environment
where $\mu_B$ (and/or $\mu_Q$) have finite values,
a strangeness symmetric distribution can be established
by adjusting $\mu_S$ appropriately: $\mu_S=\third(\mu_B-\mu_Q)$
(since $B_s=\third$, $Q_s=-\third$, and $S_s=-\third$).

It may seem odd to introduce a chemical potential
within the context of a canonical treatment,
but a given scenario may well require a canonical treatment
with respect to one attribute ({\em e.g.}\ strangeness), 
while a grand-canonical treatment suffices
with respect to another ({\em e.g.}\ baryon number).
Of course, as the above analysis brings out,
the consideration of several attributes
is relevant only when the system contains several species
that combine the attributes differently.
(The inherent correlation between baryon number and strangeness
in the quark-gluon plasma was recently proposed as a useful diagnostic
for strongly interacting matter \cite{KMR:PRL}.)

\section{General statistical treatment of strange hadrons}
\label{formulas}

We derive here the expressions needed for the general (classical) statistical 
treatment of a gas of hadrons characterized by a temperature $T$ 
and chemical potentials for baryons and electric charge,
$\mu_B$ and $\mu_Q$, with its total strangeness $S_0$ being kept fixed.
The strategy will be to group the strange hadrons species $\{\kappa\}$ 
together according to their strangeness $S_\kappa$
and then build up the complete partition function by pairwise 
inclusion of classes having opposite value of their strangeness.

\subsection{Generic multiplicities}

The one-particle partition function for a given 
strange hadronic specie $\kappa$ is given by
\beq
\zeta_\kappa(T,\mu_B,\mu_Q)\ =\
\ZERO{\zeta}_\kappa(T)\ \rme^{(\mu_B B_\kappa+\mu_Q Q_\kappa)/T}\ ,
\eeq
where its value for vanishing chemical potentials is
\beq
\ZERO{\zeta}_\kappa(T)\ =\ {g_\kappa\over2\pi^2}{VT^3\over\hbar^3c^3}
\left({m_\kappa c^2\over T}\right)^2 \!K_2\left({m_\kappa c^2\over T}\right)\ .
\eeq
The effective one-particle partition function 
for a class of hadrons having a common strangeness $S$ is then
\beqar
\zeta_S(T,\mu_B,\mu_Q) &=&
\sum_\kappa \delta_{S_\kappa,S}\, \zeta_\kappa(T,\mu_B,\mu_Q)\\ \nonumber
&=& \sum_\kappa \delta_{S_\kappa,S}\, \ZERO{\zeta}_\kappa(T)\,
\rme^{(\mu_B B_\kappa+\mu_Q Q_\kappa)/T}\ ,
\eeqar
which generally differs from $\zeta_{-S}$.
The number of hadrons having the strangeness $S$ in a given system
is denoted by $N_S$ and the probability that such a hadron belongs to 
the particular specie $\kappa$ is given by $\zeta_{\kappa}/\zeta_{S}$.
The associated first and second multiplicity moments are then
$\langle n_{\kappa'}\rangle=\langle N_{S'}\rangle\zeta_{\kappa'}/\zeta_{S'}$
and $\langle n_{\kappa'} n_{\kappa''}\rangle=
\langle N_{S'}N_{S''}\rangle\zeta_{\kappa'}\zeta_{\kappa''}/
\zeta_{S'}\zeta_{S''}$.
We may therefore concentrate on finding the generic multiplicities $\{N_S\}$.

We first combine two conjugate classes $\{+S\}$ and $\{-S\}$ 
to form the class $\{\pm S\}$.
The partition function for the resulting combined system of hadrons
having $S_\kappa=\pm S$ is then
\beqar\nonumber
\Z_{S_0}^{\{\pm S\}} &=& \sum_{N_{+S},N_{-S}}
{\zeta_{+S}^{N_{+S}}\over N_{+S}!}\, {\zeta_{-S}^{N_{-S}}\over N_{-S}!}\,
\delta_{(N_{+S}-N_{-S})S,S_0}\\
&=& \left({\zeta_{+S}\over\zeta_{-S}}\right)^{\frac{1}{2}S_0}
I_{S_0}(2\zeta_0)\ ,
\eeqar
where $N_{\pm S}\geq0$ denotes the number of hadrons
having the strangeness $S_\kappa=\pm S$
and $\zeta_0^2=\zeta_{+S}\zeta_{-S}$.
We note that $\Z_{S_0}^{\{\pm S\}}$ 
and the corresponding multiplicity moments
vanish unless $S_0$ is a multiple of $S$, 
{\em i.e.}\ $S_0=0,\pm S,\pm 2S,\dots$.
The factorial multiplicity moments can also readily be obtained,
\beqar\nonumber
&~& \langle N_{+S}(N_{+S}-1)\dots(N_{+S}-m+1)\rangle_{S_0}^{\{\pm S\}}\\
&=& \zeta_{+S}^m
\left({\zeta_{-S}\over\zeta_{+S}}\right)^{\frac{1}{2}mS}
{I_{S_0-mS}(2\zeta_0) \over I_{S_0}(2\zeta_0)}\ .
\eeqar
Furthermore, the mixed multiplicity moments
can be obtained by use of the constraint $(N_{+S}-N_{-S})S=S_0$.

These relations provide a complete treatment of one pair
of conjugate classes.
Imagine now that we have thus obtained 
the partition functions $\Z_{S_0}^{\{\pm S\}}$ 
and the corresponding multiplicity moments 
$\langle N_{\pm S}^m\rangle_{S_0}^{\{\pm S\}}$ for $|S|=1,2,3$.
The classes $\{\pm S\}$ may then be combined recursively.
Thus, combining first $\{\pm1\}$ with $\{\pm2\}$,
we find the partition function for the combined ensemble $\{\pm1,\pm2\}$,
\beq
\Z_{S_0}^{\{\pm1,\pm2\}}\
= \!\!\sum_{S_2=0,\pm2,\dots} \Z_{S_0-S_2}^{\{\pm1\}}\, \Z_{S_2}^{\{\pm2\}}\ ,
\eeq
where $S_0$ is the specified total strangeness.
It follows that if we have a self-conjugate system
({\em i.e.}\ for each hadron specie $\kappa$ included,
the corresponding antispecie $\bar\kappa$ is also included)
with strangeness $S_\kappa=\pm1,\pm2$ and whose total strangeness is $S_0$,
then the probability that those with $S_\kappa=\pm1$
have a combined strangeness of $S'$ is given by
\beq
P_{S_0}^{\{\pm1,\pm2\}}(S^{\{\pm1\}}\!=S') =
\Z_{S'}^{\{\pm1\}} \Z_{S_0-S'}^{\{\pm2\}} / \Z_{S_0}^{\{\pm1,\pm2\}} ,
\eeq
which vanishes unless $S_0-S'$ is even.
Consequently,
after the classes $\{\pm1\}$ and $\{\pm2\}$ have been combined,
the multiplicity moments for hadrons with $S_\kappa=\pm1$ are
\beqar
&~& \langle N_{S=\pm1}^m\rangle_{S_0}^{\{\pm1,\pm2\}}\ =\
\left(\Z_{S_0}^{\{\pm1,\pm2\}}\right)^{-1}\\ \nonumber
&~& \times \sum_{S'=S_0,S_0\pm2,\dots}
\Z_{S'}^{\{\pm1\}}\ \Z_{S_0-S'}^{\{\pm2\}}\ 
\langle N_{S=\pm1}^m\rangle_{S'}^{\{\pm1\}}\ ,
\eeqar
while those for hadrons with $S_\kappa=\pm2$ are
\beqar
&~& \langle N_{S=\pm2}^m\rangle_{S_0}^{\{\pm1,\pm2\}}\ =\
\left(\Z_{S_0}^{\{\pm1,\pm2\}}\right)^{-1}\\ \nonumber
&~& \times \sum_{S_2=0,\pm2,\pm4,\dots}
\Z_{S_0-S_2}^{\{\pm1\}}\ \Z_{S_2}^{\{\pm2\}}\ 
\langle N_{S=\pm2}^m\rangle_{S_2}^{\{\pm2\}}\ .
\eeqar
Similar expressions hold for the mixed moments, {\em e.g.}
\beqar
&~& \langle N_{-1}N_{+1}\rangle_{S_0}^{\{\pm1,\pm2\}}\ =\
\left(\Z_{S_0}^{\{\pm1,\pm2\}}\right)^{-1}\\ \nonumber
&~& \times \sum_{S'=S_0,S_0\pm2,\dots}
\Z_{S'}^{\{\pm1\}}\ \Z_{S_0-S'}^{\{\pm2\}}\ 
\langle N_{-1}N_{+1}\rangle_{S'}^{\{\pm1\}}\ .
\eeqar

It is straightforward to verify the following sum rule
expressing strangeness conservation,
\beq
\sum_{S=\pm1,\pm2} \langle N_{S}\rangle_{S_0}^{\{\pm1,\pm2\}} S\ =\ S_0\ .
\eeq
We note that the above recursion scheme holds even if
there are no hadrons with $S=\pm2$.
In that case, the corresponding effective partition functions vanish, 
$\zeta_{\pm2}=0$, and, as noted earlier, the combined partition function 
is unity for $S_0=0$ and vanishes otherwise,
$\Z_{S_0}^{\{\pm2\}}=\delta_{S_0,0}$.
As a consequence, the combined partition function remains unchanged
by the incorporation of $S=\pm2$,
$\Z_{S_0}^{\{\pm1,\pm2\}}=\Z_{S_0}^{\{\pm1\}}$,
and the multiplicity moments remain unchanged as well.

Further conjugate strangeness classes may be incorporated analogously.
Thus, the inclusion of $S=\pm3$ yields the following partition function
for $\{\pm1,\pm2,\pm3\}$,
\beq
\Z_{S_0}^{\{\pm1,\pm2,\pm3\}}\ =\ \sum_{S_3=0,\pm3,} 
\Z_{S_0-S_3}^{\{\pm1,\pm2\}} \Z_{S_3}^{\{\pm3\}}\ ,
\eeq
and the generic multiplicity moments are given by
\beqar
&~& \langle N_{S=\pm1}^m\rangle_{S_0}^{\{\pm1,\pm2,\pm3\}}\ =\
\left(\Z_{S_0}^{\{\pm1,\pm2,\pm3\}}\right)^{-1}\\ \nonumber  &\times&
\sum_{S'=S_0,S_0\pm3,\dots} \Z_{S'}^{\{\pm1,\pm2\}}\ \Z_{S_0-S'}^{\{\pm3\}}\ 
\langle N_{S=\pm1}^m\rangle_{S'}^{\{\pm1,\pm2\}}\ ,\\
&~& \langle N_{S=\pm2}^m\rangle_{S_0}^{\{\pm1,\pm2,\pm3\}}\ =\
\left(\Z_{S_0}^{\{\pm1,\pm2,\pm3\}}\right)^{-1}\\ \nonumber &\times&
\sum_{S'=S_0,S_0\pm3,\dots} \Z_{S'}^{\{\pm1,\pm2\}}\ \Z_{S_0-S'}^{\{\pm3\}}\
\langle N_{S=\pm2}^m\rangle_{S'}^{\{\pm1,\pm2\}}\ ,\\
&~& \langle N_{S=\pm3}^m\rangle_{S_0}^{\{\pm1,\pm2,\pm3\}}\ =\
\left(\Z_{S_0}^{\{\pm1,\pm2,\pm3\}}\right)^{-1}\\ \nonumber &\times&
\sum_{S_3=0,\pm3,\pm6,\dots}\Z_{S_0-S_3}^{\{\pm1,\pm2\}}\ \Z_{S_3}^{\{\pm3\}}\
\langle N_{S=\pm3}^m\rangle_{S_3}^{\{\pm3\}}\ .
\eeqar
and the sum rule becomes
\beq
\sum_{S=\pm1,\pm2,\pm3} 
\langle N_{S}\rangle_{S_0}^{\{\pm1,\pm2,\pm3\}} S\ =\ S_0\ .
\eeq
Recursion expressions for the correlations between the generic multiplicities
can be obtained in a similar manner.
For example, the correlations between $N_{-1}$ and $N_{+1}$ follow
from the corresponding mixed moment,
\beqar
&~& \langle N_{-1}N_{+1}\rangle_{S_0}^{\{\pm1,\pm2,\pm3\}}\ =\
\left(\Z_{S_0}^{\{\pm1,\pm2,\pm3\}}\right)^{-1}\\ \nonumber 
&~& \times\hspace{-1em}
\sum_{S'=S_0,S_0\pm3,\dots} \Z_{S'}^{\{\pm1,\pm2\}}\ \Z_{S_0-S'}^{\{\pm3\}}\ 
\langle N_{-1}N_{+1}\rangle_{S'}^{\{\pm1,\pm2\}}\ .
\eeqar
This procedure readily generalizes to the combination
of any number of self-conjugate classes.

\subsection{Individual hadron species}

The above treatment allows us to determine the canonical moments of the
generic hadron multiplicities $\{N_S\}$ 
in any blob with a specified strangeness $S_0$.
We now consider the multiplicities of the individual hadron species.

For a given value of $S_0$,
the mean number of a particular species $\kappa'$ is given by
\beq
\langle n_{\kappa'} \rangle_{S_0} = {1\over\Z_{S_0}} \prod_\kappa
\left[\sum_{n_\kappa\geq0}{\zeta_\kappa^{n_\kappa}\over n_\kappa!}\right]\!
n_{\kappa'}\, \delta(\sum_\kappa S_\kappa n_\kappa - S_0)\ ,
\eeq
and the correlation between the multiplicities of any two species 
$\kappa'$ and $\kappa''$ is given by a similar expression with
$n_{\kappa'}$ replaced by $n_{\kappa'} n_{\kappa''}$.
In order to use these expressions 
with the above expressions for the generic multiplicities $\{N_S\}$,
which no longer contain the individual species multiplicities $\{n_\kappa\}$,
we note that the probability that a generic hadron of strangeness $S=S_\kappa$
is of a particular specie $\kappa$ is given by 
$p_\kappa\equiv\zeta_\kappa/\zeta_{S_\kappa}$.
Consequently,
\beq\label{nkave}
\langle n_{\kappa'}\rangle_{S_0}\ =\ 
p_{\kappa'} \langle N_{S_{\kappa'}}\rangle_{S_0}\ .
\eeq

The second multiplicity moments are more complicated to obtain,
because they receive contributions both from the correlated internal
multiplicity fluctuations within the separate $S$ classes
characterized by each particular generic multiplicity set $\{N_S\}$
and from the correlated fluctuations of these generic multiplicities.
It is convenient to introduce the correlation coefficient
$\sigma_{\kappa'\kappa''}\equiv
\delta_{S'S''}\,p_{\kappa'}(\delta_{\kappa'\kappa''}-p_{\kappa''})$
which expresses the degree of correlation between two particles
in the same strangeness class.
Thus it vanishes if $S'\neq S''$.
When $S'=S''$ we have $\sigma_{\kappa'\kappa''}=-p_{\kappa'}p_{\kappa''}$ 
when the two species differ,
$\kappa'\neq\kappa''$, while $\sigma_{\kappa'\kappa''}=p_{\kappa'}q_{\kappa'}$ 
when $\kappa'=\kappa''$,
where $p_{\kappa'}\equiv\zeta_{\kappa'}/\zeta_{S'}$ is the probability
that a hadron of the class $S'$ belongs to the specie $\kappa'$
and $q_{\kappa'}\equiv1-p_{\kappa'}$ is its complement.
We then find
\beq\label{ave-nn}
\langle n_{\kappa'}n_{\kappa''}\rangle_{S_0}
= p_{\kappa'} p_{\kappa''} \langle N_{S'}N_{S''}\rangle_{S_0} + 
\delta_{S'S''}\langle N_{S'}\rangle_{S_0}\sigma_{\kappa'\kappa''}\ .
\eeq
The multiplicity covariances then involve both 
the intra-class correlations $\sigma_{N'N''}$
and the inter-class correlations $\sigma_{n'n''}=
\delta_{S'S''}\langle N'\rangle_{S_0}\sigma_{\kappa'\kappa''}$,
\beqar\label{sig-nn}
\sigma_{n_{\kappa'}n_{\kappa''}}^{S_0}
&\equiv& \langle n_{\kappa'}n_{\kappa''}\rangle_{S_0}
-\langle n_{\kappa'}\rangle_{S_0} \langle n_{\kappa''} \rangle_{S_0}\\ 
\nonumber &=& p_{\kappa'} p_{\kappa''}\, \sigma_{N'N''}\ 
+\ \delta_{S'S''}\, \langle N'\rangle_{S_0}\sigma_{\kappa'\kappa''}\ .
\eeqar

The above expressions allow us to evaluate the average multiplicities
of individual hadron species as well as the associated (co)variances,
for any given value of the fixed strangeness of the blob, $S_0$.

\section{Global canonical treatment}
\label{global}

We consider here the more complicated situation
where $N$ individual blobs are subject to a global canonical constraint
on their combined strangeness, $S_{\rm tot}\equiv\sum_n S_n$.
We specialize to the relevant case of $S_{\rm tot}\!=\!0$,
but the treatment can readily be adapted to any value.

Assuming, as we have throughout, that all the blobs are similar,
the joint probability for finding the combined system with 
$\nu_n$ $s$ quarks and $\bar{\nu}_n$ $\bar{s}$ antiquarks
in the blob $n$ is then  given by
\beqar
&~& P(\nu_1,\bar{\nu}_1;\dots;\nu_N\bar{\nu}_N)\ =\\ \nonumber
 &~& {1\over I_0(2N\zeta_s)}
{\zeta_s^{\nu_1}\over\nu_1!}{\zeta_s^{\bar{\nu}_1}\over\bar{\nu}_1!}\cdots
{\zeta_s^{\nu_N}\over\nu_N!}{\zeta_s^{\bar{\nu}_N}\over\bar{\nu}_N!}\
\delta_{\nu_1+\dots+\nu_N,\bar{\nu}_1\dots+\bar{\nu}_N}\ .
\eeqar
This expression can be used as a basis for a direct Metropolis sampling 
of the individual multiplicities $\{\nu_n,\bar{\nu}_n\}$,
starting for example from $\{0,0\}$.
However, this may not be optimally efficient,
since in fact we only need to know the distribution of the differences
$S_n=\bar{\nu}_n-\nu_n$, which is given by
\beqar\nonumber
&~& P(S_1,\dots,S_N)\\  \nonumber
&=& \prod_{n=1}^N \left[\sum_{\nu_n\bar{\nu}_n}
\delta_{\bar{\nu}_n-\nu_n,S_n}\right]
P(\nu_1,\bar{\nu}_1;\dots;\nu_N\bar{\nu}_N)\\ \nonumber
&=& {\delta_{S_1+\dots+S_N,0}\over I_0(2N\zeta_s)} \prod_{n=1}^N
\left[\sum_{\nu_n\bar{\nu}_n}
{\zeta_s^{\nu_n+\bar{\nu}_n}\over\nu_n!\bar{\nu}_n!}
\delta_{\bar{\nu}_n-\nu_n,S_n}\right]\\
&=& {\delta_{S_1+\dots+S_N,0}\over I_0(2N\zeta_s)}
\prod_{n=1}^N\left[ I_{S_n}(2\zeta_s)\right]\ .
\eeqar
The corresponding Metropolis sampling could then start from $\{S_n\}=\{0\}$,
for example, and repeatedly exchange one unit of strangeness between two
selected subsystems, $S_i\to S_i'=S_i\pm1$ and $S_j\to S_j'=S_j\mp1$,
with the corresponding ratio of weights being given in terms of ratios
of Bessel functions of neighboring order,
$W'/W=
(I_{S_i\pm1}(2\zeta_s)/I_{S_i}(2\zeta_s))
(I_{S_j\mp1}(2\zeta_s)/I_{S_j}(2\zeta_s))$.

A much simpler procedure is based on the fact that
the entire set of configurations $\{\nu_n,\bar{\nu}_n\}$
can be organized according to the total number of $s$ quarks present, $M$
(which equals the total number of $\bar s$ when $S_{\rm tot}\!=\!0$).
The class having $M=0$ contains only the empty configuration $\{0,0\}$,
while the class having $M=1$ has $N^2$ members, and so on.
Thus, class $M$ has $(N^M/M!)^2$ members,
each of which has the relative weight $\zeta_s^{2M}$,
and it is readily checked that the sum of weights is $I_0(2N\zeta_s)$,
the total partition fucntion.
The expected number of $s$ quarks is $N\zeta_s$
and the most likely value of $M$ is $[N\zeta_s]$.
It is easy to sample $M$ by a Metropolis procedure 
based on the weight ratios for adjacent values of $M$,
${W_{M+1}/ W_M} = ({N\zeta_s/ M+1})^2$ and 
${W_{M-1}/ W_M} = ({M/ N\zeta_s})^2$.

Alternatively, $M$ could be sampled directly from 
its analytical distribution $P(M)=\zeta_s^{2M}(N^M/M!)^2/I_0(2N\zeta_s)$ 
in a standard manner.
Once $M$ has been selected, it is straightforward to distribute
the $M$ quarks and $M$ antiquarks randomly among the $N$ subsystems 
and thus obtain their strangenesses as $S_n=\bar{\nu}_n-\nu_n$.

The above discussion brings out the fact that
the total plasma partition function can be calculated
by either distributing $\nu_n$ quarks and $\bar{\nu}_n$ antiquarks 
in each subsystem $n$ or distributing $M$ quarks and $\bar M$ antiquarks
anywhere within the combined system,
\beqar\label{Q0}
&~& \prod_{n=1}^N \left[\sum_{\nu_n\bar{\nu}_n}
{\zeta_s^{\nu_n}\over\nu_n!}{\zeta_s^{\bar{\nu}_n}\over\bar{\nu}_n!}\right]
\delta_{\nu_1+\ldots+\nu_N,\bar{\nu}_1+\ldots+\bar{\nu}_N}\\ \nonumber
&=& \sum_{M\bar M}{(N\zeta_s)^M\over M!}{(N\zeta_s)^{\bar M}\over\bar{M}!}\
\delta_{M,\bar M}\ =\ I_0(2N\zeta_s)\ .
\eeqar

In any case,
once the individual blob strangenesses $\{S_n\}$ have been selected,
the various hadron multiplicities can be sampled as in the standard case
discussed in Appendix \ref{formulas}.
It is thus possible, in this manner, to make a complete statistical
sampling of the combined system and then perform any analysis of interest.

However, such a full similation is not always necessary.
In particular, as is often the case,
when the quantities of interest can be expressed
in terms of first and second moments of the multiplicity distributions 
of specified hadron species,
it is possible to employ recursion relations 
in analogy with those derived in Appendix \ref{formulas} for
$\{\langle n^{(i)}_\kappa\rangle_{S_i}\}$,
the average multiplicities of the hadron species $\kappa$
emitted by the particular blob $i$,
and the corresponding second moments,
$\{\langle n^{(i)}_\kappa n^{(i)}_{\kappa'}\rangle_{S_i}\}$.
(This approach can of course be extended to higher moments.)

We first note that the canonical partition function for $N$ blobs
can be expressed recursively in terms of those for fewer blobs.
For example, with $N=N'+(N-N')$,
\beq
I_S(2N\zeta_s)\ =\ \sum_{S'} I_{S'}(2N'\zeta_s)\ I_{S-S'}(2(N-N')\zeta_s)\ .
\eeq
The inclusive probability for one particular blob to have a given strangeness
can then be expressed as follows,
\beqar\nonumber  
&~& P_1^{(1\cdots N)}(S_1)\ =\ \sum_{S_2\dots S_N} P(S_1,\dots,S_N)\\
&~& =\ {I_{S_1}(2\zeta_s)\ I_{-S_1}(2(N-1)\zeta_s) \over I_0(2N\zeta_s)}\ ,
\eeqar
while the inclusive joint probability for two given blobs
to have specified strangenesses is of the following form,
\beqar\nonumber  
&~& P_{12}^{(1\cdots N)}(S_1,S_2)\ =\ 
\sum_{S_3\dots S_N} P(S_1,S_2,\dots,S_N)\\
&~& =\ {I_{S_1}(2\zeta_s)\ I_{S_2}(2\zeta_s)\ I_{-S_1-S_2}(2(N-2)\zeta_s) 
\over I_0(2N\zeta_s)}\ ,
\eeqar

It is then straightforward to express the average multiplicity 
of the hadron species $\kappa$ arising from a given blob,
\beq
\langle n^{(1)}_\kappa\rangle_0^{(1\cdots N)}\ =\
\sum_{S_1} P_1^{(1\cdots N)}(S_1) \langle n^{(1)}_\kappa\rangle_{S_1}\ .
\eeq
A similar expression holds for the average of any power of that multiplicity, 
$(n^{(1)}_\kappa)^m$.
It also readily follows that the mixed multiplicity moment for 
two hadron species emitted from the same blob is
\beq
\langle n^{(1)}_\kappa n^{(1)}_{\kappa'} \rangle_0^{(1\cdots N)} =
\sum_{S_1} P_1^{(1\cdots N)}(S_1)
\langle n^{(1)}_\kappa n^{(1)}_{\kappa'} \rangle_{S_1}\ ,
\eeq
while the corresponding expression for emission from two different blobs is
\beq
\langle n^{(1)}_\kappa n^{(2)}_{\kappa'} \rangle_0^{(1\cdots N)} =
\sum_{S_1S_2} P_{12}^{(1\cdots N)}(S_1,S_2)
\langle n^{(1)}_\kappa \rangle_{S_1} \langle n^{(2)}_{\kappa'} \rangle_{S_2}\ .
\eeq
These expressions allow us to use the canonical partition functions 
for individual blobs obtained in Appendix \ref{formulas}
to calculate the average multiplicities
of specific hadron species and any desired (co)variances in terms of the
corresponding expressions for canonical emission from a single blob.

\section*{Acknowledgments}

\hspace*{\parindent}
A.M.\ would like to thank F.\ Beccatini for helpful discussions.
This work was supported by the Director, Office of Energy Research,
Office of High Energy and Nuclear Physics, Division of Nuclear Physics,
the Office of Basic Energy
Science, Division of Nuclear Science, of the U.S. Department of Energy under
Contract No. DE-AC03-76SF00098.

{}

			\end{document}